# Quantum flux effects on the energy spectra and thermo-magnetic properties in 2D Schrodinger equation with Mobius square potential

A.N.Ikot[1], U.S.Okorie[2], I.B.Okon[3], P.O.Amadi[1], N.Okpara[4], L.F.Obagboye[5],

A.I.Ahmadov[6,7], H.Horchani[8], A.-H. Abdel-Aty[9] and C.Duque[10]

[1]Theoretical Physics Group, Department of Physics, University of Port Harcourt, Choba, Nigeria.
[2]Department of Physics, Akwa Ibom State University, Ikot Akpaden, P.M.B. 1167, Uyo, Nigeria.
[3]Theoretical Physics Group, Department of Physics, University of Uyo, Nigeria.
4 Delta State University, Abraka, Delta state-Nigeria
[5]National Mathematical Centre, Abuja, Nigeria.
[6]Department of Theoretical Physics, Baku State University, Z. Khalilov st. 23, AZ1148, Baku, Azerbaijan.
[7]Institute for Physical Problems, Baku State University, Z. Khalilov st. 23, AZ1148, Baku, Azerbaijan.
[8]Department of Physics, College of Science, Sultan Qaboos University, P.O. Box 36, P. C. 123, Al-Khod, Muscat, Sultanate of Oman.
.
[9]Department of Physics, College of Sciences, University of Bisha, P. O. Box 344, Bisha 61922, Saudi Arabia
[10] Grupo de Materia Condensada-UdeA, Instituto de Física, Facultad de Ciencias Exactas y Naturales, Universidad de Antioquia UdeA, Calle 70 No. 52-21, Medellín, Colombia

## Abstract

A 2D Schrodinger equation with interacting Mobius square potential model is solved using Nikiforov-Uvarov Functional Analysis (NUFA) formalism. The energy spectra and the corresponding wave function for the linearly and exponentially varying quantum magnetic flux are obtained analytically in a closed form. The evaluated energy spectra are used to obtain an expression for the partition functions for the two cases comprises of the linearly and exponentially varying quantum magnetic flux and vis-a-vis is use to evaluate other thermodynamic and magnetic properties for the system. The results are used to study the free energy, mean energy, the entropy, specific heat, magnetization, magnetic susceptibility and the persistent current of the system. The numerical bound state energies are computed.

# 1 Introduction

In recent years the study of two dimensional Schrodinger equation [1-3] has attracted the attention of many researchers in the fields because of its vast applications in quantum dot [4-5], quantum wells [6-7]. Thermodynamic properties [8-10] and optical properties [11]. One of the investigated potential models in physics is Mobius square potential [12-13] where other potential models are embedded as special cases. Mobius potential has many applications in different branches of physics including molecular physics, chemical physics and high energy physics [14]. The Mobius square potential is defined as [12-13],

$$V(\rho) = V_0 \left( \frac{a + be^{-\alpha\rho}}{q + \eta e^{-\alpha\rho}} \right) \qquad (1)$$

where $V_0$ is the potential depth, $\alpha$ is the screening parameter, $a, b, q$ and $\eta$ are all potential parameters. The Mobius square potential of equation (1) reduced to the well-known potential models such as Morse potential, Hulthen potential, Deng-Fan potential, Morse potential, Manning-Rosen potential, Poschl-Teller potential, Tietz potential .Eckart potential, Hua potential and attractive potential as special cases. In addition, because of the complicated nature of the Mobius square potential many work had not been carried out with this potential. A great numbers of analytical techniques have been proposed and developed for solving Schrodinger, Klein-Gordon and Dirac equation with various phenomenological potentials, have been investigated which include asymptotic iteration method (AIM) [15], the Nikiforov-Uvarov (NU) technique [16], supersymmetric quantum mechanics (SUSSYQM) [17], shape invariance (SI) [18], quantization rule [19] and Nikiforov Uvarov Functional Analysis (NUFA) method [20]. Mobius square potential have been investigated within the framework of spin and pseudospin symmetry[21]. In addition different potential models have been investigated within relativistic and non- relativistic quantum mechanics[22]. The Schrodinger equation in the present of external magnetic and AB fields are usually referred to as 2D-Schrodinger equation is defined as [1-3],

$$\left( i\hbar\vec{\nabla} - \frac{e}{c}\vec{A} \right)^2 \psi(\rho,\varphi,z) = 2\mu[E + V(\rho)]\psi(\rho,\varphi,z) \qquad (2)$$

where $\mu$ is the reduced mass, $E$ is the energy level of the system, $\vec{A}$ is the external magnetic vector field, $e$ and $c$ represent the electronic charge and the speed of light respectively. The two-dimensional nonrelativistic equation was obtained from the Schrodinger equation by incorporating the external magnetic fields. This was done by replacing the momentum as $\vec{P} \to \left(\vec{P} - \frac{e}{c}\vec{A}\right)$ where $\vec{P} \to i\hbar\nabla$ is the momentum operator, Horchani et al.[1] studied 2D Schrodinger equation with inversely square potential. Several authors have devoted their time to investigate the effects of magnetic and Aharonov–Bohm flux fields on quantum systems[2]. Ikot et al.[3] studied the Schrodinger equation with screened Kratzer potential in the present of external magnetic fields. The interband transitions in quantum pseudodot system in the present of external magnetic field have been examined [23]. Aygun et al.[24] used AIM to study the 2D Schrodinger equation with Kratzer potential. Very recently, Ikot et al.[25] investigated the 2D Schrodinger with new Morse interacting potential. A 2D Klein–Gordon equation with harmonic oscillator has also investigated. The 2D Pauli equation with Hulthen potential for spin-1/2 particle in the presence of Aharonov–Bohm (AB) field had been studied by Ferkous and Bounames[26] Many research work have been carried out for various potential functions to study the thermodynamic properties of different systems, using different techniques [27-30] and Euler–MacLaurin [30-31].

The purpose of this research work is in threefold. Firstly, we will find the solution of equation (2) with the potential given in equation (1) using the recently proposed Nikiforov-Uvarov Functional Analysis (NUFA) method [20]. Secondly, we will analysed obtained energy spectra linearly and exponential depending on the behaviour of the quantum flux. Thirdly, we will investigate the thermodynamic properties for the system.

## 3 Determination of energy spectra in 2D Schrodinger equation with Mobius Square potential

The 2D Schrodinger equation with Mobius Square interacting potential is written as,

$$\left\{\frac{1}{2\mu}\left(i\hbar\vec{\nabla} - \frac{e}{c}\vec{A}\right)^2\right\}\psi(\rho,\varphi) = (E_{nm} - V(\rho))\psi(\rho,\varphi), \qquad (3)$$

where $E_{nm}$ denotes the energy level, $\mu$ is the effective mass of the system and $\vec{A}$ is the vector potential. The details derivation and relationship between the external magnetic field B and the AB fields are given in refs.[1-3]. However, the vector potential in this study takes the form

$$\vec{A} = \vec{A}_1 + \vec{A}_2 = \left(0, \frac{Be^{-\alpha\rho}}{(q + \eta e^{-\alpha\rho})} + \frac{\phi_{AB}}{2\pi\rho}, 0\right). \qquad (4)$$

In order to solve equation (3) we adopt the following approaches:

(i) Express the Laplacian operator in cylindrical coordinate as,

$$\nabla^2 = \frac{1}{\rho}\frac{\partial}{\partial \rho}\left(\rho\frac{\partial}{\partial \rho}\right) + \frac{1}{\rho^2}\frac{\partial^2}{\partial \varphi^2} + \frac{\partial^2}{\partial z^2} \tag{5}$$

(ii) Approximate the centrifugal barrier as [33]

$$\rho^{-2} \approx \lim_{\alpha \to 0}\left\{\frac{\alpha^2 q^2}{\left(q+\eta e^{-\alpha \rho}\right)^2}\right\} \approx \lim_{\alpha \to 0}\begin{pmatrix} \frac{1}{r^2} + \frac{\alpha}{r} + \frac{5}{12}\alpha^2 + \frac{1}{12}\alpha^3\rho + \frac{1}{240}\alpha^4\rho^2 \\ -\frac{1}{720}\alpha^5\rho^3 - \frac{1}{6045}\alpha^6\rho^4 + O(\rho^5) \end{pmatrix} \tag{6}$$

(iii) Used ansatz for the wave function in the form

$$\psi(\rho,\varphi) = \frac{1}{\sqrt{2\pi\rho}} e^{im\varphi} R_{nm}(\rho) \tag{7}$$

(iv) take the angular momentum $\vec{L}$ as, $\vec{L} = -i\hbar\nabla, \nabla = \frac{1}{\rho}\frac{\partial}{\partial \varphi}$, where $m$ denotes the magnetic quantum number. Using equation (3) together with conditions (i)-(iii) and the coordinate transformation $\chi = -\frac{\eta}{q}e^{-\alpha\rho}$, then equation (3) becomes,

$$\frac{d^2 R_{nm}(\chi)}{d\chi^2} + \frac{1-\chi}{\chi(1-\chi)}\frac{dR_{nm}(\chi)}{d\chi}$$
$$+ \frac{1}{\chi^2(1-\chi)^2}\left\{\begin{array}{l}-\left(\frac{\varepsilon_{n,m}^2}{\alpha^2} + \left(\frac{\tau_0}{\eta}\right)^2 + \frac{2\mu V_0 b^2}{\hbar^2 \eta^2}\right)\chi^2 + \left(\frac{2\varepsilon_{n,m}^2}{\alpha^2} + \frac{\tau q}{\eta} + \frac{2\mu V_0}{\hbar^2 \eta q}\right)\chi \\ -\left(\frac{\varepsilon_{n,m}^2}{\alpha^2} + (m+\varsigma)^2 - \frac{1}{4} + \frac{2\mu V_0 a^2}{\hbar^2 q^2}\right)\end{array}\right\} R_{n,m}(\chi) = 0 \tag{8}$$

where, $\varsigma = \frac{\phi_{AB}}{\phi_0}$ is an integer and the flux quantum is defined as $\phi_0 = \frac{hc}{e}, \tau_0 = \frac{eB}{\hbar c}$ and $\tau$ is given in linear and exponential form as follows (a) $\tau = \tau_0\left(1 + \frac{\varsigma}{2}\right)$, (b) $\tau = \tau_0 \exp\left(\frac{\varsigma}{2}\right)$. The plots of the linearly and exponentially quantum flux is shown in Fig.1. Figure 1 shows clearly that the form of the quantum flux will have considerable effects on the behavior of the quantum system. These effects will be examine on the energy spectra and the thermodynamic properties of the system when it is either linearly or exponentially as the case maybe.

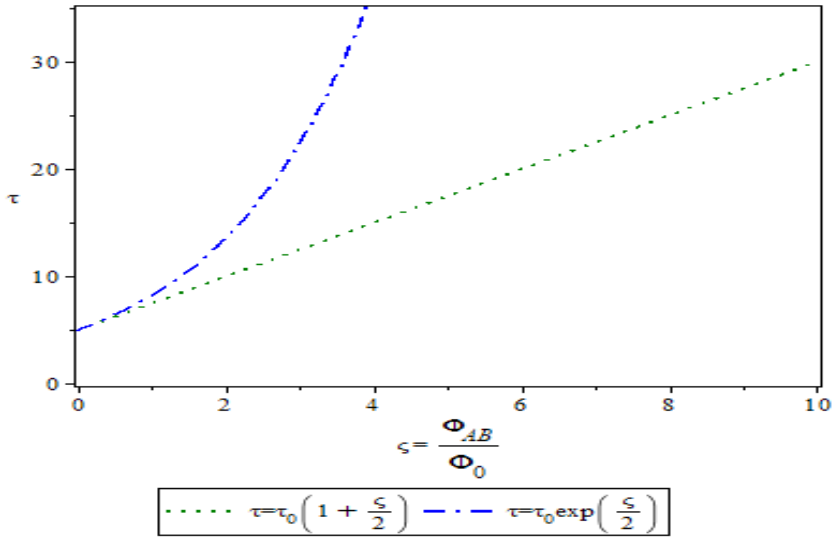

Figure 1. The plots of the linearly and exponentially quantum flux

Equation (8) can be simplifies as follows:

$$\frac{d^2 R_{nm}(\chi)}{d\chi^2} + \frac{1-\chi}{\chi(1-\chi)}\frac{dR_{nm}(\chi)}{d\chi} + \frac{1}{\chi^2(1-\chi)^2}\left\{-\xi_1\chi^2 + \xi_2\chi - \xi_3\right\}R_{n,m}(\chi) = 0 \qquad (9)$$

where,

$$\xi_1 = \left(\frac{\varepsilon_{n,m}^2}{\alpha^2} + \left(\frac{\tau_0}{\eta}\right)^2 + \frac{2\mu V_0 b^2}{\hbar^2 \eta^2}\right), \xi_2 = \left(\frac{2\varepsilon_{n,m}^2}{\alpha^2} + \frac{\tau q}{\eta} + \frac{2\mu V_0}{\hbar^2 \eta q}\right),$$

$$\xi_3 = \left(\frac{\varepsilon_{n,m}^2}{\alpha^2} + \left((m+\varsigma)^2 - \frac{1}{4}\right) + \frac{2\mu V_0 a^2}{\hbar^2 q^2}\right) \qquad (10)$$

Using equation (A8) and (A9) of the NUFA formulation, we get

$$\lambda = \sqrt{(m+\varsigma)^2 - \frac{1}{4} + \frac{2\mu V_0 a^2}{\hbar^2 q^2} + \frac{\varepsilon^2}{\alpha^2}}$$

$$v = \frac{1}{2} \pm \frac{1}{2}\sqrt{1 + 4\left(\frac{\tau_0^2}{\eta^2} - \frac{q\tau}{\eta} + \frac{2\mu V_0}{\hbar}\left(\frac{b^2}{\eta^2} + \frac{a^2}{q^2} - \frac{1}{q\eta}\right) + (m+\varsigma)^2 - \frac{1}{4}\right)} \qquad (11)$$

Substituting equation (11) into equation (A11) gives the energy spectra for the 2D Schrödinger equation as,

$$E_{n,m} = -\frac{\hbar^2 \alpha^2}{2\mu}\left(\frac{\Xi}{2(n+v)} + \frac{n+v}{2}\right)^2 + \frac{\hbar^2 \alpha^2}{2\mu}\left((m+\varsigma)^2 - \frac{1}{4} + \frac{2\mu V_0 a^2}{\hbar^2 q^2}\right) \qquad (12)$$

where,

$$\Xi = (m+\varsigma)^2 - \frac{1}{4} + \frac{2\mu V_0}{\hbar^2}\left(\frac{a^2}{q^2} - \frac{b^2}{\eta^2}\right) - \frac{\tau_0^2}{\eta^2} \tag{13}$$

The corresponding wave function for the system is obtain as follows,

$$\psi(s) = N\left(-\frac{\eta}{q}e^{-\alpha\rho}\right)^{\sqrt{\frac{\varepsilon_{n,m}^2}{\alpha^2}+(m+\varsigma)^2-\frac{1}{4}+\frac{2\mu V_0 a^2}{\hbar^2 q^2}}}\left(1+\frac{\eta}{q}e^{-\alpha\rho}\right)^{\frac{1}{2}+\frac{1}{2}\sqrt{1+4\left(\frac{\tau_0^2}{\eta^2}-\frac{q\tau}{\eta}+\frac{2\mu V_0}{\hbar}\left(\frac{b^2}{\eta^2}+\frac{a^2}{q^2}-\frac{1}{q\eta}\right)+(m+\varsigma)^2-\frac{1}{4}\right)}} {}_2F_1(\delta,g,\sigma;s) \tag{14}$$

where $\delta$, $g$, $\sigma$ are given as follows,

$$\delta = \begin{pmatrix} \sqrt{(m+\varsigma)^2 - \frac{1}{4} + \frac{2\mu V_0 a^2}{\hbar^2 q^2} + \frac{\varepsilon^2}{\alpha^2} + \frac{1}{2}} \\ +\frac{1}{2}\sqrt{1+4\left(\frac{\tau_0^2}{\eta^2} - \frac{q\tau}{\eta} + \frac{2\mu V_0}{\hbar}\left(\frac{b^2}{\eta^2} + \frac{a^2}{q^2} - \frac{1}{q\eta}\right) + (m+\varsigma)^2 - \frac{1}{4}\right)} \\ +\sqrt{\frac{\varepsilon_{n,m}^2}{\alpha^2} + \left(\frac{\tau_0}{\eta}\right)^2 + \frac{2\mu V_0 b^2}{\hbar^2 \eta^2}} \end{pmatrix} \tag{15}$$

$$g = \begin{pmatrix} \sqrt{(m+\varsigma)^2 - \frac{1}{4} + \frac{2\mu V_0 a^2}{\hbar^2 q^2} + \frac{\varepsilon^2}{\alpha^2} + \frac{1}{2}} \\ +\frac{1}{2}\sqrt{1+4\left(\frac{\tau_0^2}{\eta^2} - \frac{q\tau}{\eta} + \frac{2\mu V_0}{\hbar}\left(\frac{b^2}{\eta^2} + \frac{a^2}{q^2} - \frac{1}{q\eta}\right) + (m+\varsigma)^2 - \frac{1}{4}\right)} \\ -\sqrt{\frac{\varepsilon_{n,m}^2}{\alpha^2} + \left(\frac{\tau_0}{\eta}\right)^2 + \frac{2\mu V_0 b^2}{\hbar^2 \eta^2}} \end{pmatrix} \tag{16}$$

$$\sigma = 1 + 2\sqrt{(m+\varsigma)^2 - \frac{1}{4} + \frac{2\mu V_0 a^2}{\hbar^2 q^2} + \frac{\varepsilon^2}{\alpha^2}} \tag{17}$$

## 4 Determination of thermo-magnetic properties for Mobius Square

### 4.1 With linearly varying quantum magnetic flux $\tau = \tau_0\left(1+\frac{\Phi_{AB}}{2\Phi_0}\right)$ Using Poisson summation

In this section we will investigate the thermodynamic properties of the system when the quantum magnetic flux is varying linearly. It is well-known that all thermodynamic properties can be obtained from the partition function of the system [34-35]. This vibrational partition function can be computed by straightforward summation over all possible vibrational energy levels accessible to the system. The partition function $Z(\beta)$ at finite temperature $T$ is obtained with the Boltzmann factor as [34-35];

$$Z(\beta) = \sum_{n=0}^{\infty} e^{-\beta E_n} \tag{18}$$

with $\beta = \dfrac{1}{kT}$ and with $k$ is the Boltzmann constant. All thermodynamic and magnetic properties such as the free energy, mean energy, the entropy, specific heat, magnetization, magnetic susceptibility and the persistent current, can be obtained from the partition function $Z(\beta)$. These thermodynamic functions for the diatomic molecules system can be calculated from the following expressions [7-9];

$$F(\beta) = -\frac{1}{\beta}\ln Z(\beta),$$

$$U(\beta) = -\frac{d\ln Z(\beta)}{d\beta},$$

$$S(\beta) = \ln Z(\beta) - \beta\frac{d\ln Z(\beta)}{d\beta},$$

$$C(\beta) = \beta^2 \frac{d^2 \ln Z(\beta)}{d\beta^2}, \tag{19}$$

$$M(\beta) = \frac{1}{\beta}\left(\frac{1}{Z(\beta)}\right)\left(\frac{\partial}{\partial \vec{B}}Z(\beta)\right),$$

$$\chi_m(\beta) = \frac{\partial M(\beta)}{\partial \vec{B}},$$

$$I(\beta) = -\frac{e}{hc}\frac{\partial F(\beta)}{\partial m}$$

Energy spectrum defined in equation (12) can be simplifed as

$$E_{nm} = -p\left[\frac{\Xi}{2(n+\nu)} + \frac{n+\nu}{2}\right]^2 + Q \tag{20}$$

where

$$p = \frac{\hbar^2 \alpha^2}{2\mu}, \quad Q = \frac{\hbar^2 \alpha^2}{2\mu}\left[(m+\varsigma)^2 - \frac{1}{4} + \frac{2\mu V_0 a^2}{\hbar^2 q^2}\right]^2, \tag{21}$$

The partition function $Z(\beta, \lambda)$ at finite temperature $T$ is obtained with the Boltzman factor in this form:

$$Z(\beta,\lambda) = \sum_{n=0}^{\lambda} e^{-\beta E_{nm}}, \quad \beta = \frac{1}{k_B T} \tag{22}$$

Substituting equation (20) into (22) gives

$$Z(\beta,\lambda) = e^{-pQ} \sum_{n=0}^{\lambda} e^{\beta p \left[\frac{\Xi}{2(n+\nu)} + \frac{n+\nu}{2}\right]^2} \qquad (23)$$

In evaluating Eq.(23), we employ the Poisson summation formula for lower order approximation as

$$\sum_{n=0}^{n_{max}} f(n) = \frac{1}{2}[f(0) - f(n_{max}+1)] + \int_0^{n_{max}+1} f(x)dx \qquad (24)$$

Expressing equation (23) Using equation (24) gives the partition function as

$$Z(\beta,\lambda) = \frac{1}{2}\left[e^{\beta p g_1^2} - e^{\beta p g_2^2}\right]e^{-\beta Q} + e^{-\beta Q}\int_0^{\lambda+1} e^{\beta p \left(\frac{\Xi}{2(x+\nu)} + \frac{x+\nu}{2}\right)^2} dx \qquad (25)$$

where

$$g_1 = \frac{\Xi}{2\nu} + \frac{\nu}{2}; \quad g_2 = \frac{\Xi}{2(\lambda+\nu+1)} + \frac{\lambda+\nu+1}{2}. \qquad (26)$$

However,

$$\int_0^{\lambda+1} f(x)dx = e^{-\beta Q}\int_0^{\lambda+1} e^{\beta p \left(\frac{\Xi}{2(x+\nu)} + \frac{x+\nu}{2}\right)^2} dx \qquad (27)$$

We introduce new variable $z = \frac{\Xi}{2(x+\nu)} + \frac{x+\nu}{2}$ this then implies

$$(x+\nu)^2 - 2z(x+\nu) + \Xi = 0 \qquad (28)$$

$$x + \nu = z \pm \sqrt{z^2 - \Xi} \Rightarrow x = z \pm \sqrt{z^2 - \Xi} - \nu$$

where

$$x \geq 0, \ x = z + \sqrt{z^2 - \Xi} - \nu \qquad (29)$$

Expressing equation (27) in terms of (26) gives

$$\int_0^{\lambda+1} f(x)dx = e^{-\beta Q}\int_0^{\lambda+1} e^{\beta p \left(\frac{\Xi}{2(x+\nu)} + \frac{x+\nu}{2}\right)^2} dx = e^{-\beta Q}\int_{g_1}^{g_2} e^{\beta p z^2}\left(1 + \frac{z}{\sqrt{z^2-\Xi}}\right)dz =$$

$$= e^{-\beta Q}\int_{g_1}^{g_2} e^{\beta p z^2} dz + e^{-\beta Q}\int_{g_1}^{g_2} \frac{z e^{\beta p z^2} dz}{\sqrt{z^2-\Xi}} = e^{-\beta Q}\frac{1}{2}\sqrt{\frac{\pi}{\beta p}}\left[erfi(\sqrt{\beta p}\cdot g_2) - erfi(\sqrt{\beta p}\cdot g_1)\right] +$$

$$+e^{-\beta Q}\frac{1}{2}\sqrt{\frac{\pi}{\beta p}}\left[erfi(\sqrt{\beta p(g_2^2-\Xi)}))-erfi(\sqrt{\beta p(g_1^2-\Xi)}))\right] \qquad (30)$$

Let

$$I_1=\int_{g_1}^{g_2}e^{\beta p z^2}dz=\frac{1}{2}\sqrt{\frac{\pi}{\beta p}}\left[erfi(\sqrt{\beta p}\cdot g_2)-erfi(\sqrt{\beta p}\cdot g_1)\right] \qquad (31)$$

$$I_2=\int_{g_1}^{g_2}\frac{ze^{\beta p z^2}dz}{\sqrt{z^2-\Xi}}=\int_{g_1}^{g_2}ze^{\beta p z^2}d\sqrt{z^2-\Xi}=\left\{\begin{array}{l}\tau=\sqrt{\beta p(z^2-\Xi)}\\ d\tau=\sqrt{\beta p}d(z^2-\Xi)\end{array}\right\}$$
$$=\frac{1}{\sqrt{\beta p}}\int_{\sqrt{\beta p(g_1^2-\Xi)}}^{\sqrt{\beta p(g_2^2-\Xi)}}e^{\tau^2}d\tau \Rightarrow \frac{1}{2}\sqrt{\frac{\pi}{\beta p}}\left[erfi(\sqrt{\beta p(g_2^2-\Xi)})-erfi(\sqrt{\beta p(g_1^2-\Xi)}))\right] \qquad (32)$$

where $erfi(u)=\frac{2}{\sqrt{\pi}}\int_0^u e^{\tau^2}d\tau$ is the imaginary error function.

Therefore, the thermodynamic and thermomagnetic properties are obtained as follows:

**I Partition function**

$$Z(\beta,\lambda)=\frac{1}{2}\left[e^{\beta(pg_1^2-Q)}-e^{\beta(pg_2^2-Q)}\right]+\frac{1}{2}\sqrt{\frac{\pi}{\beta p}}e^{-\beta Q}\left[\begin{array}{l}erfi(\sqrt{\beta p}\cdot g_2)+erfi(\sqrt{\beta(pg_2^2-\Xi)})\\ -[erfi(\sqrt{\beta p}\cdot g_1)-erfi(\sqrt{\beta(pg_1^2-\Xi)})]\end{array}\right] \qquad (33)$$

**II Vibrational internal energy**

$$U(\beta,\lambda)=-\frac{\partial}{\partial \beta}(\ln Z(p,\lambda))=-\frac{1}{Z(\beta,\lambda)}\frac{\partial Z(p,\lambda)}{\partial \beta}=$$
$$=-\frac{1}{Z(\beta,\lambda)}\frac{1}{2}e^{-\beta Q}\left\{(pg_1^2-Q-\frac{g_1}{\beta})e^{\beta pg_1^2}-(pg_2^2-Q-\frac{g_2}{\beta})e^{\beta pg_2^2}+\frac{1}{\beta}\left[\sqrt{g_2^2-\Xi}e^{\beta p(g_2^2-\Xi)}-\sqrt{g_1^2-\Xi}e^{\beta p(g_1^2-\Xi)}\right]\right. \qquad (34)$$
$$\left.-\left(Q+\frac{1}{2\beta}\right)\sqrt{\frac{\pi}{\beta p}}e^{-\beta Q}\left[erfi(\sqrt{\beta p}g_2)+erfi(\sqrt{\beta p(g_2^2-\Xi)})-erfi(\sqrt{\beta p}g_1)-erfi(\sqrt{\beta p(g_1^2-\Xi)})\right]\right\}$$

## III Vibrational free energy

$$F(\beta,\lambda) = -\frac{1}{\beta}\ln Z(\beta,\lambda)$$

$$= -\frac{1}{\beta}\ln\left\{\frac{1}{2}\left[e^{\beta(pg_1^2-Q)} - e^{\beta(pg_2^2-Q)}\right] + \frac{1}{2}\sqrt{\frac{\pi}{\beta p}}e^{-\beta Q}\begin{bmatrix}\text{erfi}(\sqrt{\beta p}\cdot g_2) + \text{erfi}(\sqrt{\beta(pg_2^2-\Xi)}) \\ -[\text{erfi}(\sqrt{\beta p}\cdot g_1) - \text{erfi}(\sqrt{\beta(pg_1^2-\Xi)})]\end{bmatrix}\right\} \quad (35)$$

## IV Vibrational entropy

$$S(\beta,\lambda) = -k_B \ln Z(\beta,\lambda) - k_B \beta \frac{\partial}{\partial \beta}\ln Z(\beta,\lambda) = k_B \ln Z(\beta,\lambda) -$$

$$-\frac{k_B\beta}{Z(\beta,\lambda)}\frac{1}{2}e^{-\beta Q}\left\{\begin{aligned}&(pg_1^2-Q-\frac{g_1}{\beta})e^{\beta pg_1^2} \\ &-(pg_2^2-Q-\frac{g_2}{\beta})e^{\beta pg_2^2}\frac{1}{\beta}\left[\sqrt{g_2^2-\Xi}\,e^{\beta p(g_2^2-\Xi)} - \sqrt{g_1^2-\Xi}\,e^{\beta p(g_1^2-\Xi)}\right]\left(Q+\frac{1}{2\beta}\right) \\ &\sqrt{\frac{\pi}{\beta p}}\begin{bmatrix}\text{erfi}(\sqrt{\beta p}\,g_2) + \text{erfi}(\sqrt{\beta p(g_2^2-\Xi)}) \\ -\text{erfi}(\sqrt{\beta p}\,g_1) - \text{erfi}(\sqrt{\beta p(g_1^2-\Xi)})\end{bmatrix}\end{aligned}\right\} \quad (36)$$

## V Vibrational specific heat capacity

$$C(\beta,\lambda) = k_B\beta^2 \frac{\partial^2}{\partial\beta^2}\ln Z(\beta,\lambda) = k_B\beta^2 \frac{\partial}{\partial\beta}\left[\frac{\partial}{\partial\beta}\ln Z(\beta,\lambda)\right]$$

$$= k_B\beta^2\left\{\frac{-1}{Z^2(\beta,\lambda)}\frac{1}{4}e^{-2\beta Q}\left[(pg_1^2-Q-\frac{g_1}{\beta})e^{\beta pg_1^2} - (pg_2^2-Q-\frac{g_2}{\beta})e^{\beta pg_2^2}\right.\right.$$

$$+\frac{1}{\beta}\left[\sqrt{g_2^2-\Xi}\,e^{\beta p(g_2^2-\Xi)} - \sqrt{g_1^2-\Xi}\,e^{\beta p(g_1^2-\Xi)}\right]$$

$$\left.-\left(Q+\frac{1}{2\beta}\right)\sqrt{\frac{\pi}{\beta p}}\left[\text{erfi}(\sqrt{\beta p}\,g_2) + \text{erfi}(\sqrt{\beta p(g_2^2-\Xi)}) - \text{erfi}(\sqrt{\beta p}\,g_1) - \text{erfi}(\sqrt{\beta p(g_1^2-\Xi)})\right]^2\right.$$

$$-\frac{Qe^{\beta Q}}{2Z(\beta,\lambda)}\left[(pg_1^2-Q-\frac{g_1}{\beta})e^{\beta pg_1^2} - (pg_2^2-Q-\frac{g_2}{\beta})e^{\beta pg_2^2}\right]$$

$$\left.+\frac{1}{\beta}\left[\sqrt{g_2^2-\Xi}\,e^{\beta p(g_2^2-\Xi)} - \sqrt{g_1^2-\Xi}\,e^{\beta p(g_1^2-\Xi)}\right] - \left(Q+\frac{1}{2\beta}\right)\sqrt{\frac{\pi}{\beta p}}\begin{bmatrix}\text{erfi}(\sqrt{\beta p}\,g_2) + \text{erfi}(\sqrt{\beta p(g_2^2-\Xi)}) \\ -\text{erfi}(\sqrt{\beta p}\,g_1) - \text{erfi}(\sqrt{\beta p(g_1^2-\Xi)})\end{bmatrix}\right\}$$

$$+\frac{e^{\beta Q}}{2Z(\beta,\lambda)}\cdot\begin{bmatrix}(pg_1^2-Q-\frac{g_1}{\beta})pg_1^2+\frac{g_1}{\beta})e^{\beta pg_1^2}-(pg_2^2-Q-\frac{g_2}{\beta})pg_2^2+\frac{g_2}{\beta})e^{\beta pg_2^2}\\-\frac{1}{\beta^2}\left[\sqrt{g_2^2-\Xi}e^{\beta p(g_2^2-\Xi)}-\sqrt{g_1^2-\Xi}e^{\beta p(g_1^2-\Xi)}\right]+\frac{1}{\beta}\left[p(g_2^2-\Xi)e^{\beta p(g_2^2-\Xi)}-p(g_1^2-\Xi)e^{\beta p(g_1^2-\Xi)}\right]\\+\frac{1}{2\beta}\sqrt{\frac{\pi}{\beta p}}\left(Q+\frac{3}{2\beta}\right)\left[erfi(\sqrt{\beta p}g_2)+erfi(\sqrt{\beta p(g_2^2-\Xi)})-erfi(\sqrt{\beta p}g_1)+erfi(\sqrt{\beta p(g_2^2-\Xi)})\right]\\-\left(Q+\frac{1}{2\beta}\right)\frac{1}{\beta}\left[g_2e^{\beta pg_2^2}+\sqrt{g_2^2-\Xi}e^{\beta p(g_2^2-\Xi)}g_1e^{\beta pg_1^2}+\sqrt{g_1^2-\Xi}e^{\beta p(g_1^2-\Xi)}\right]\end{bmatrix} \quad (37)$$

## VI Persistent Current

$$I(\beta)=-\frac{e}{\hbar c}\frac{\partial F(\beta)}{\partial m}$$

$$I(\beta,\lambda)=-\frac{e}{\hbar c}\frac{\partial}{\partial m}F(\beta,\lambda)=-\frac{e}{\hbar c}\frac{\partial}{\partial m}\left[-\frac{1}{\beta}\ln Z(\beta,\lambda)\right]=\frac{e}{\beta\hbar c}\frac{1}{Z(\beta,\lambda)}\frac{\partial}{\partial m}Z(\beta,\lambda)=$$

$$=\frac{e}{\hbar c\beta Z(\beta,\lambda)}\frac{\partial}{\partial m}\left\{\frac{1}{2}\left[e^{\beta(pg_1^2-Q)}-e^{\beta(pg_2^2-Q)}\right]+\right.$$

$$\left.+\frac{1}{2}\sqrt{\frac{\pi}{\beta p}}e^{-\beta Q}\left[erfi(\sqrt{\beta p}g_2)+erfi(\sqrt{\beta p(g_2^2-\Xi)})-erfi(\sqrt{\beta p}g_1)-erfi(\sqrt{\beta p(g_1^2-\Xi)})\right]\right\}=$$

$$=\frac{e}{\hbar c\beta Z(\beta,\lambda)}\left\{\frac{1}{2}\left[\beta\left(2pg_1\frac{\partial g_1}{\partial m}-\frac{\partial Q}{\partial m}\right)e^{\beta(pg_1^2-Q)}-\beta\left(2pg_2\frac{\partial g_2}{\partial m}-\frac{\partial Q}{\partial m}\right)e^{\beta(pg_2^2-Q)}\right]\right.$$

$$-\frac{\beta}{2}\sqrt{\frac{\pi}{\beta p}}e^{-\beta Q}\frac{\partial Q}{\partial m}\left[erfi(\sqrt{\beta p}g_2)+erfi(\sqrt{\beta p(g_2^2-\Xi)})-erfi(\sqrt{\beta p}g_1)-erfi(\sqrt{\beta p(g_1^2-\Xi)})\right] \quad (38)$$

$$+\frac{1}{2}\sqrt{\frac{\pi}{\beta p}}e^{-\beta Q}\begin{bmatrix}\frac{2}{\sqrt{\pi}}\sqrt{\beta p}\frac{\partial g_2}{\partial m}e^{\beta pg_2^2}+\frac{2}{\sqrt{\pi}}\sqrt{\beta p}\frac{2g_2\frac{\partial g_2}{\partial m}-\frac{\partial\Xi}{\partial m}}{2\sqrt{g_2^2-\Xi}}e^{\beta p(g_2^2-\Xi)}-\frac{2}{\sqrt{\pi}}\sqrt{\beta p}\frac{\partial g_1}{\partial m}e^{\beta pg_1^2}\\-\frac{2}{\sqrt{\pi}}\sqrt{\beta p}\frac{2g_1\frac{\partial g_1}{\partial m}-\frac{\partial\Xi}{\partial m}}{2\sqrt{g_1^2-\Xi}}e^{\beta p(g_1^2-\Xi)}\end{bmatrix}$$

## VII Magnetic Susceptibility

$$\chi_m(\beta) = \frac{\partial M(\beta)}{\partial B}$$

$$\chi_m(\beta,\lambda) = \frac{\partial M(\beta,\lambda)}{\partial B} = \frac{\partial}{\partial B}\frac{e^{-\beta Q}}{\beta Z(\beta,\lambda)}\left\{\left(\beta p g_1 - 1 - \frac{g_1 e^{-\beta p \Xi}}{\sqrt{g_1^2 - \Xi}}\right)\right.$$

$$e^{\beta p g_1^2}\left[-\frac{e\tau_0}{\hbar c \eta^2 \nu} + \left(\frac{1}{2} - \frac{\Xi}{2\nu^2}\right)\frac{e}{\hbar c(2\nu - 1)}\left(\frac{2\tau_0}{\eta^2} - \frac{g}{2}\left(1 + \frac{\varsigma}{2}\right)\right)\right] +$$

$$+\left(-\beta p g_2 + 1 + \frac{g_2 e^{-\beta p \Xi}}{\sqrt{g_2^2 - \Xi}}\right)e^{\beta p g_2^2} \cdot \frac{e\tau_0}{\hbar c \eta^2}\left[-\frac{1}{\nu + \lambda + 1} + \frac{1}{2\sqrt{-\Xi}} + \frac{\sqrt{-\Xi}}{2(\nu + \lambda + 1)}\right] -$$

$$e^{-\beta p \Xi}\left(\frac{e^{\beta p g_1^2}}{\sqrt{g_1^2 - \Xi}} - \frac{e^{\beta p g_2^2}}{\sqrt{g_2^2 - \Xi}}\right)\frac{e\tau_0}{\hbar c \eta^2}\right\} \quad (39)$$

**VIII Magnetization**

$$M(\beta,\lambda) = \frac{1}{\beta}\frac{1}{Z(\beta,\lambda)}\frac{\partial Z(\beta,\lambda)}{\partial B} = M(\beta,\lambda) = \frac{1}{\beta}\frac{1}{Z(\beta,\lambda)}\frac{\partial}{\partial B}\left\{\frac{1}{2}\left[e^{\beta(pg_1^2-Q)} - e^{\beta(pg_2^2-Q)}\right] + \frac{1}{2}\sqrt{\frac{\pi}{\beta p}}e^{-\beta Q}\right.$$

$$\left[erfi(\sqrt{\beta p} \cdot g_2) + erfi\left(\sqrt{\beta p(g_2^2 - \Xi)}\right) - erfi(\sqrt{\beta p} \cdot g_1)\right.$$
$$\left.\left. - erfi\left(\sqrt{\beta p(g_1^2 - \Xi)}\right)\right]\right\} = \frac{1}{\beta}\frac{1}{Z(\beta,\lambda)}\left\{\frac{1}{2}\beta p 2 \cdot g_1 \frac{dg_1}{dB}e^{\beta(pg_1^2 - Q)}\right.$$

$$-\frac{1}{2}\beta p 2 g_2 \frac{dg_2}{dB}e^{\beta(pg_2^2 - Q)} + \frac{1}{2}\sqrt{\frac{\pi}{\beta p}}e^{-\beta Q}\left[\frac{2}{\sqrt{\pi}}\sqrt{\beta p}\frac{dg_2}{dB}e^{\beta p g_2^2} + \frac{2}{\sqrt{\pi}}\sqrt{\beta p}\frac{1}{2\sqrt{g_1^2 - \Xi}} \cdot \right.$$

$$\left(2g_2 \frac{dg_2}{dB} - \frac{d\Xi}{dB}\right)e^{\beta p(g_2^2 - \Xi)} - \frac{2}{\sqrt{\pi}}\sqrt{\beta p}\frac{dg_1}{dB}e^{\beta p g_1^2} - \frac{2}{\sqrt{\pi}}\sqrt{\beta p}\frac{1}{2\sqrt{g_1^2 - \Xi}}\left(2g_1\frac{dg_1}{dB} - \right.$$

$$\left.\left.\frac{d\Xi}{dB}\right)e^{\beta p(g_1^2 - \Xi)}\right]\right\} = \frac{1}{\beta}\frac{1}{Z(\beta,\lambda)} \cdot \left\{\beta p g_1 e^{\beta(pg_1^2 - Q)}\frac{dg_1}{dB} - \beta p g_2 e^{\beta(pg_2^2 - Q)}\frac{dg_2}{dB} + \right.$$

$$e^{-\beta Q}\left[e^{\beta p g_1^2}\frac{dg_2}{dB}+\frac{1}{2\sqrt{g_2^2-\Xi}}\left(2g_2\frac{dg_2}{dB}-\frac{d\Xi}{dB}\right)e^{\beta p(g_2^2-\Xi)}-e^{\beta p g_1^2}\frac{dg_1}{dB}-\frac{1}{2\sqrt{g_1^2-\Xi}}\left(2g_1\frac{dg_1}{dB}-\right.\right.$$
$$\left.\left.\frac{d\Xi}{dB}\right)e^{\beta p(g_1^2-\Xi)}\right]\right\} \qquad (40)$$

## 4.2 With exponentially varying quantum magnetic flux $\tau = \tau_0 \exp\left(\frac{\varsigma}{2}\right)$.

In this section we shall determine the thermomagnetic properties with exponentially varying quantum magnetic flux. Using the same Poisson's summation concept and with the help of Mathematica software 10.0 version, the partition function for exponentially varying quantum magnetic flux is given as

$$Z(\beta)=e^{-\beta Q}\left\{\frac{2\left(e^{\beta Q_1 P_1^2}-e^{\beta Q_1 P_2^2}\right)\sqrt{\beta}\sqrt{Q_1}+\sqrt{\pi}\left[\begin{array}{c}erf\sqrt{\beta}\sqrt{Q_1}P_2-erf\sqrt{\beta}\sqrt{Q_1}P_2\\ +e^{4\beta Q_1\Xi}\left(\begin{array}{c}erf\sqrt{\beta}\sqrt{Q_1}\sqrt{P_1^2-4\Xi}\\ -erf\sqrt{\beta}\sqrt{Q_1}\sqrt{P_2^2-4\Xi}\end{array}\right)\end{array}\right]}{4\sqrt{\beta}\sqrt{Q_1}}\right\}. \qquad (41)$$

Using equation equation (41), other thermomagnetic properties for exponentially varying quantum magnetic flux can be obtained.

## 5.0 Results and Discussion

Table 1 is the numerical Bound state solution under the influence of AB and Magnetic field for fixed magnetic for linearly varying quantum magnetic flux using physical constant parameter $a=b=1, V_0=0.2, \alpha=0.2, q=0.1, \eta=0.3$. In table 1, degeneracies occurs when the Magnetic flux and Aharanov-Bohm flux field is absence ($\varsigma=\tau_0=0$). The degeneracy is completely removed with the combined presence of magnetic and Ahranov-Bohm flux field ($\varsigma=\tau_0=0.2$). Also, the numerical bound state energies decreases with an increase in quantum state for all magnetic quantum spin. Table 2 is the numerical bound state solution under the influence of AB and Magnetic field for exponentially varying quantum magnetic flux. $a=b=1, V_0=0.2, \alpha=0.2, q=0.1, \eta=0.3$. degeneracies occurs for some magnetic quantum spin. The numerical bound state energies also decreases with an increase in quantum state. Figure 1(a-d) are the variations of Partition function with varying linear quantum flux with respect to Magnetic flux ($\tau_0$), Aharanov-Bohm flux ($\zeta$), inverse temperature parameter and maximum vibrational quantum number ($\lambda$) respectively. In figure 1(a) the partition function increases exponentially with quantized spacing in increasing value of magnetic flux. In figure 1(b) the partition decreases exponentially and converges at $\zeta = 10$. In figure 1(c), the partition function decreases exponentially in a converging manner. Finally in figure 1(d), the partition function increases monotonically with an increase in maximum vibrational quantum number.

Figure 2 (a-d) are the variations of vibrational mean energy with varying linear quantum flux with respect to Magnetic flux ($\tau_0$), Aharanov-Bohm flux ($\zeta$), inverse temperature parameter and maximum vibrational quantum number ($\lambda$) respectively. In figure 2(a), the vibrational mean energy increases exponentially with an increase in magnetic flux and decreases with an increase in AB flux. In figure 2(c), the vibrational mean energy increases monotonically with an increase in inverse temperature parameter. However in figure 2(d), the vibrational mean energy is a quantized linear graph with unique equal spacing that increases with an increase in maximum vibrational quantum number. Figure 3(a-d) are the variations of vibrational heat capacity with varying linear quantum flux with respect to Magnetic flux ($\tau_0$), Aharanov-Bohm flux ($\zeta$), inverse temperature parameter and maximum vibrational quantum number ($\lambda$) respectively. In figure 3(a), the vibrational heat capacity decreases exponentially with an increasing value of the magnetic flux. In figure 3(b), the vibrational heat capacity depict a laminar flow that decreases from the vertical axis and converge at $\zeta = 122$. Also in figures 3(c) and 3(d) the vibrational heat capacity all increases monotonically with an increase in inverse temperature parameter and maximum vibrational quantum number respectively. Figure 4(a-d) are the variations of vibration entropy with varying linear quantum flux with respect to Magnetic flux ($\tau_0$), Aharanov-Bohm flux ($\zeta$), inverse temperature parameter and maximum vibrational quantum number ($\lambda$) respectively. In figure 4(a), the vibrational entropy is a montonic graph that interwoven at $\tau_0 = 7\ and\ 16$ .Tesla respectively. While in figures 4(c) and 4(d) the vibrational entropy all increases exponentially with an increase in $\beta$ and $\lambda$ respectively.

Figure 5(a-d) are the variations of vibrational free energy with varying linear quantum flux with respect to Magnetic flux ($\tau_0$), Aharanov-Bohm flux ($\zeta$), inverse temperature parameter and maximum vibrational quantum number ($\lambda$) respectively. In figure 5(a), the vibrational free energy is exponential graph that decreases with an increase in magnetic flux. In figure 5(b), the vibrational free energy is a linear quantized graph with unique spacing that increases with increasing value of Aharanov-Bohm flux ($\zeta$). In figure 5(c), the vibrational free energy increases

exponentially while in figure 5(d), the vibrational free energy has different vertical spectral lines with respect to maximum vibrational quantum number. Figure 6(a-d) are the variations of persistent current with varying linear quantum flux with respect to Magnetic flux ($\tau_0$), Aharanov-Bohm flux ($\zeta$), inverse temperature parameter and maximum vibrational quantum number ($\lambda$) respectively. In figure 6(a), the persistent current is an exponential graph that is quantized uniquely with equal spacing in an increasing value of magnetic flux. In figure 6(b), the persistent current starts from the positive y-axis and decreases with an increasing value of AB flux. However, in figure 6(c) and 6(d), the persistent current all increases monotonically with and increasing value of $\beta$ and $\lambda$ respectively.

Figure 7(a-d) are the variations of magnetic susceptibility with varying linear quantum flux with respect to Magnetic flux ($\tau_0$), Aharanov-Bohm flux ($\zeta$), inverse temperature parameter and maximum vibrational quantum number ($\lambda$) respectively. In figure 7(a), the magnetic susceptibility increases exponentially with an increase in magnetic flux. However, in figures 7(b), 7(c) and 7(d), the magnetic susceptibility all increases linearly with an increase in AB flux, inverse temperature parameter $\beta$ and maximum vibrational quantum number ($\lambda$) respectively. Figure 8(a-d) are the variations of magnetization with varying linear quantum flux with respect to Magnetic flux ($\tau_0$), Aharanov-Bohm flux ($\zeta$), inverse temperature parameter and maximum vibrational quantum number ($\lambda$) respectively. In figure 8(a), the magnetization decreases exponentially with an increase in magnetic flux before converging at $\tau_0 = 14.5$ Tesla. In figures 8(b) and 8(c) magnetization increases exponentially in an increasing value of AB field and inverse temperature parameter ($\beta$) respectively. Figure 9(a-d) are variations of partition function with varying exponential quantum flux with respect to magnetic flux ($\tau_0$), Aharanov-Bohm flux ($\zeta$), inverse temperature parameter and maximum vibrational quantum number ($\lambda$) respectively. In figure 9(a), the partition function has various local maximum turning points and increases with an increasing value of magnetic flux. In figure 9(b), the partition function starts from the positive y-axis and decreases nonlinearly before converging at $\zeta = 8.0$ at In figure 9(c), the partition function decreases exponentially with increasing value of $\beta$. In figure 9(d), the partition function is a linear graph that increases with increasing value of maximum vibrational quantum number ($\lambda$).

Figure 10(a-d) are variations of vibrational mean energy with varying exponential quantum flux with respect to magnetic flux ($\tau_0$), Aharanov-Bohm flux ($\zeta$), inverse temperature parameter and maximum vibrational quantum number ($\lambda$) respectively. In figure 10(a), the vibrational mean energy is a parabolic curve that has moderate turning point which converges at $\tau_0 = 8$ Tesla. Correspondingly, the vibrational mean

energy increases exponentially with respect to Aharanov-Bohm flux ($\zeta$) as shown in figure 10(b), monotonically with respect to $\beta$ as shown in figure 10(c) and appears linearly with conspicuous spacing with respect maximum vibrational quantum number as shown in figure 10(d). Figure 11(a-d) are variations of vibrational heat capacity with varying exponential quantum flux with respect to magnetic flux ($\tau_0$), Aharanov-Bohm flux ($\zeta$), inverse temperature parameter and maximum vibrational quantum number ($\lambda$) respectively. In figure 11(a), the vibrational heat capacity is a quantized graph that converges when the magnetic field $\tau_0 = 3.0$. Tesla. In figures 11(b), 11(c) and 11(d). The vibrational heat capacity all increases exponentially with an increasing value of $\zeta, \beta$ and $\lambda$. Respectively. Figures 12(a-d) are variations of vibrational entropy with varying exponential quantum flux with respect to magnetic flux ($\tau_0$), Aharanov-Bohm flux ($\zeta$), inverse temperature parameter and maximum vibrational quantum number ($\lambda$) respectively. In figure 12(a), the vibrational entropy has various local maximum turning points which increases with the magnetic flux. In figure 12(b), the entropy decreases from the positive y-axis and increases nonlinearly before converging at $\zeta = 8$. In figure 12(c), the entropy decreases exponentially with increasing value of $\beta$. While in figures 12(d) the entropy is a quantized linear graph with large spacing which increases with increasing value of $\lambda$.

Figures 13(a-d) are variations of vibrational free energy with varying exponential quantum flux with respect to magnetic flux ($\tau_0$), Aharanov-Bohm flux ($\zeta$), inverse temperature parameter and maximum vibrational quantum number ($\lambda$) respectively. In figure 13(a), the vibrational free energy is a parabolic curve with various minimum turning points that increases with an increasing value of magnetic field. In figure 13(b), the vibrational free energy increases asymptotically with an increasing value of AB field. In figure 13 (c), the vibrational mean energy increases monotonically increasing value of $\beta$ while in figure 13(d), the vibrational free energy is a nonlinear quantized graph with unequal spacing. Figures 14(a-d) are variations of magnetisation with varying exponential quantum flux with respect to magnetic flux ($\tau_0$), Aharanov-Bohm flux ($\zeta$), inverse temperature parameter and maximum vibrational quantum number ($\lambda$) respectively. In figures 14(a) and 14(d), the magnetization is a straight graph that increases with increasing value of magnetic field and maximum vibrational quantum number respectively. In figure 14(b), magnetization showcases parabolic curve with various distinct maximum turning point with a sharp peak at $\zeta = 0.03$. In figure 14(c), magnetization is a nonlinear graph that decreases monotonically with increasing value of maximum vibrational quantum number. Figures 15(a-d) are variations of magnetic susceptibility with varying exponential quantum flux with respect to magnetic flux ($\tau_0$), Aharanov-

Bohm flux ($\varsigma$), inverse temperature parameter and maximum vibrational quantum number ($\lambda$) respectively. In figure 15(a), the magnetic susceptibility is a half sinusoidal graph with distinct maximum turning points which slopes and converges at $\tau_0 = 14.5$ Tesla. In figure 15(b), the magnetic susceptibility is a quantized graph with local maximum turning point. In figure 15(c), the magnetic susceptibility is a spectral graph that start from the origin before splitting into different unique quantized curves which increases with increasing value of inverse temperature parameter. In figure 15(d), the magnetic susceptibility increases monotonically with an increasing value of maximum vibrational quantum number. Figures 16(a-d) are variations of persistent current with varying exponential quantum flux with respect to magnetic flux ($\tau_0$), Aharanov-Bohm flux ($\varsigma$), inverse temperature parameter and maximum vibrational quantum number ($\lambda$) respectively. In figure 16(a), the persistent current is a linear graph with unique intersection at $\tau_0 = 1.5$ Tesla. In figures 16(b), the persistent current increases exponentially with increasing value of $\varsigma$. In figures 16(c) and 16(d) the persistent current increases monotonically with increasing value of $\beta$ and $\lambda$ respectively.

Table 1: Numerical Bound state solution under the influence of AB and Magnetic field for fixed magnetic for linearly varying quantum magnetic flux using $a = b = 1, V_0 = 0.2, \alpha = 0.2, q = 0.1, \eta = 0.3$

| $m$ | $n$ | $(\varsigma = \tau_0 = 0)eV$ | $(\varsigma = 0.2, \tau_0 = 0)eV$ | $(\varsigma = 0, \tau_0 = 0.2)eV$ | $(\varsigma = \tau_0 = 0.2)eV$ |
|---|---|---|---|---|---|
| 0 | 0 | 0.4140573410 | 0.4179027610 | 0.4638591595 | 0.4179027610 |
|   | 1 | 0.4002780881 | 0.4044674868 | 0.4290782462 | 0.4044674868 |
|   | 2 | 0.3274177382 | 0.3316421869 | 0.3316421869 | 0.33164221869 |
|   | 3 | 0.2186552190 | 0.2227726040 | 0.2325659850 | 0.2227726040 |
| -1 | 0 | 0.5097806661 | 0.4754132366 | 0.5625261810 | 0.4754132366 |
|   | 1 | 0.5046881182 | 0.4671794688 | 0.5363338664 | 0.4671794688 |
|   | 2 | 0.4332678480 | 0.3951183852 | 0.4553470018 | 0.3951183852 |
|   | 3 | 0.3224585080 | 0.2849070470 | 0.3387106345 | 0.2849070470 |
| 1 | 0 | 0.5097806661 | 0.5516616974 | 0.5625261810 | 0.5516616974 |
| | 1 | 0.5046881182 | 0.5504113561 | 0.5363338664 | 0.5504113561 |
| | 2 | 0.4332678480 | 0.4799236060 | 0.4553470018 | 0.4799236060 |
| | 3 | 0.3224585080 | 0.3685592470 | 0.3387106345 | 0.3685592470 |

Table 2: Numerical Bound state solution under the influence of AB and Magnetic field for fixed magnetic for exponentially varying quantum magnetic flux. $a=b=1, V_0=0.2, \alpha=0.2, q=0.1, \eta=0.3$

| $m$ | $n$ | $(\varsigma=\tau_0=0)eV$ | $(\varsigma=0.2, \tau_0=0)eV$ | $(\varsigma=0, \tau_0=0.2)eV$ | $(\varsigma=\tau_0=0.2)eV$ |
|---|---|---|---|---|---|
| 0 | 0 | -0.3359426590 | -0.3400972390 | -0.2861408405 | -0.2901633745 |
| | 1 | -0.3497219119 | -0.3535325132 | -0.3209217538 | -0.3244871792 |
| | 2 | -0.4225822618 | -0.4263578131 | -0.4029864378 | -0.4064647625 |
| | 3 | -0.5313447810 | -0.5352273960 | -0.5174340150 | -0.5209693725 |
| -1 | 0 | -0.4402193339 | -0.4025867634 | -0.3874738190 | -0.3508142471 |
| | 1 | -0.4453118818 | -0.4108205312 | -0.4136661336 | -0.3800163108 |
| | 2 | -0.5167321520 | -0.4828816148 | -0.4946529982 | -0.4614574695 |
| | 3 | -0.6275414920 | -0.5930929530 | -0.6112893655 | -0.5773899055 |
| 1 | 0 | -0.4402193339 | -0.4863383026 | -0.3874738190 | -0.2901633745 |
| | 1 | -0.4453118818 | -0.4875886439 | -0.4136661336 | -0.3244871792 |
| | 2 | -0.5167321520 | -0.5580763940 | -0.4946529982 | -0.4064647625 |
| | 3 | -0.6275414920 | -0.6694407530 | -0.6112893655 | -0.5209693725 |

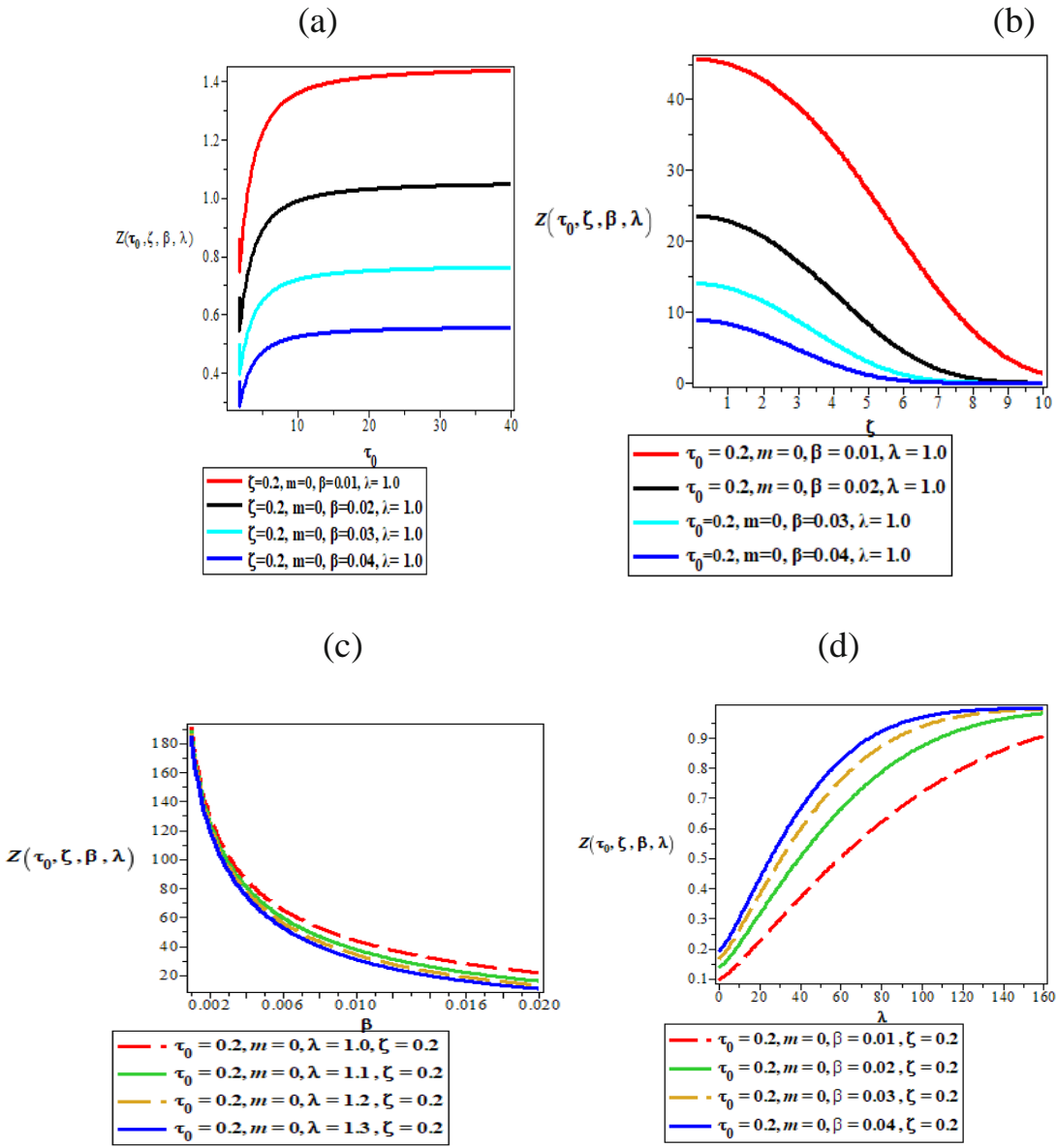

Figure 1. Variation of Partition function with varying linear quantum flux with respect to (a) Magnetic flux ($\tau_0$), (b) Aharanov-Bohm flux ($\zeta$) (c) Inverse temperature parameter and (d) Maximum vibrational quantum number ($\lambda$).

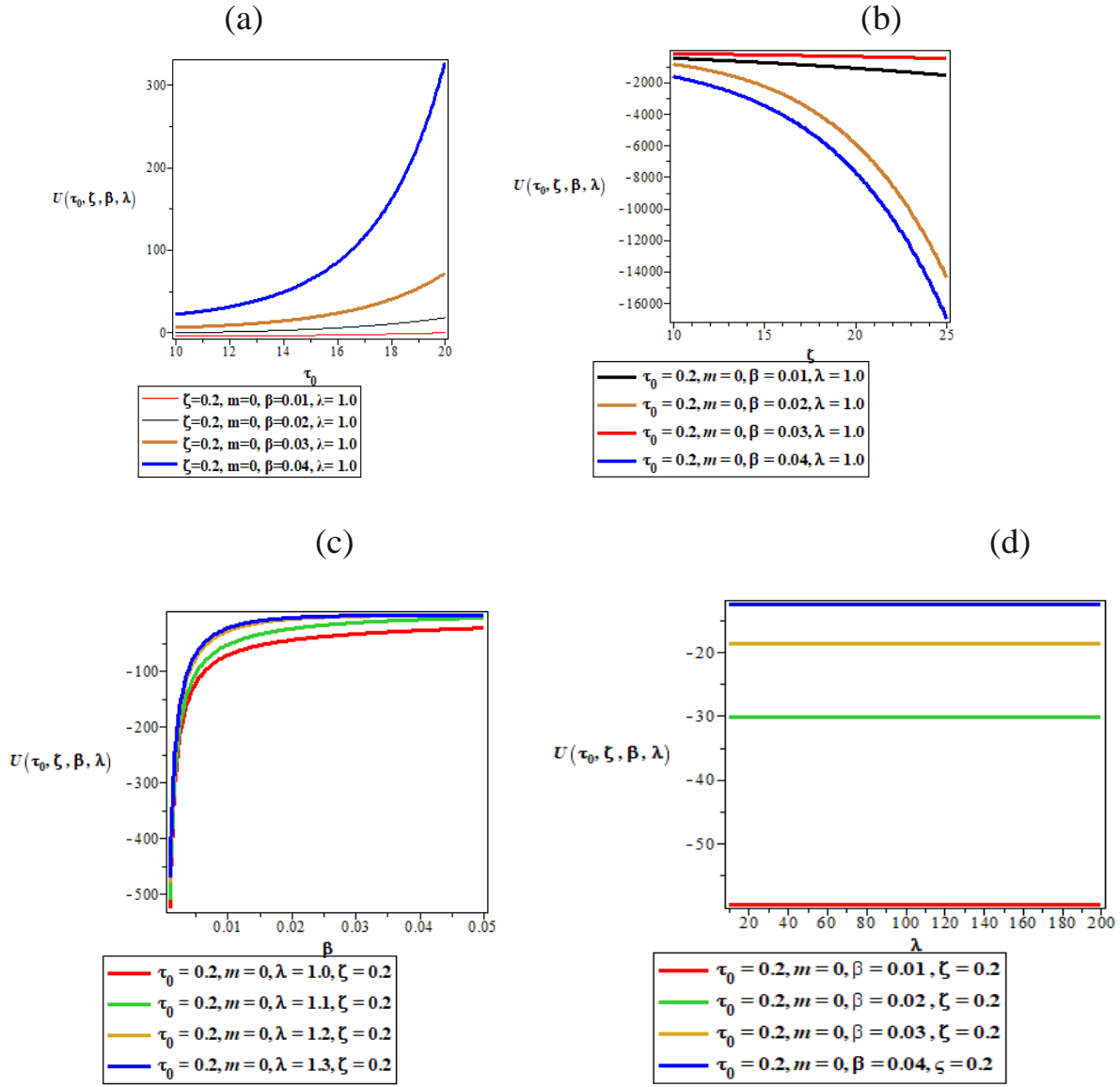

Figure 2: Variation of maximum vibrational energy with varying linear quantum flux with respect to (a) Magnetic flux ($\tau_0$), (b) Aharanov-Bohm flux ($\zeta$) (c) Inverse temperature parameter and (d) Maximum vibrational quantum number ($\lambda$).

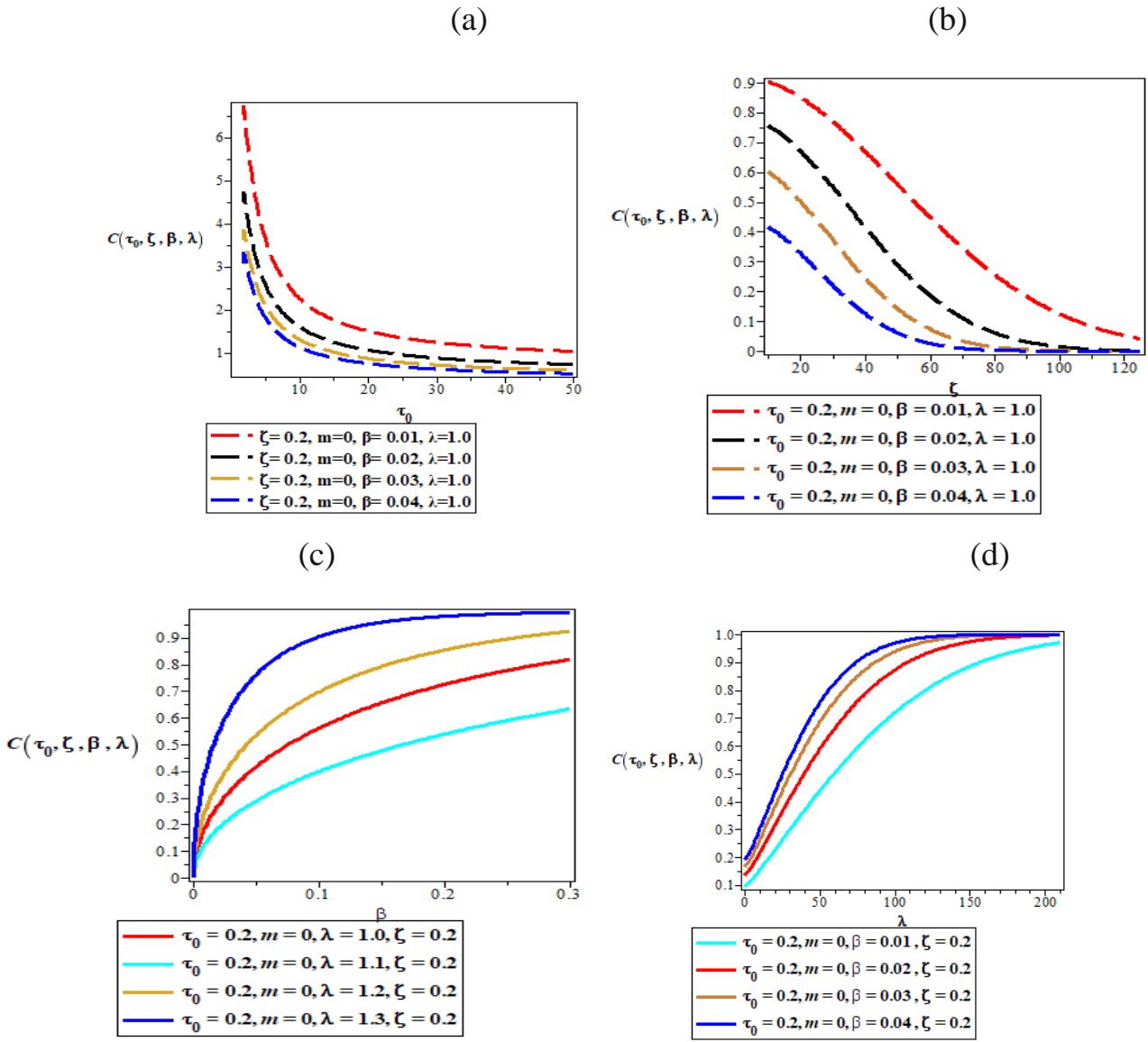

Figure 3: Variation of vibrational heat capacity with varying linear quantum flux with respect to (a) Magnetic flux ($\tau_0$), (b) Aharanov-Bohm flux ($\zeta$) (c) Inverse temperature parameter and (d) Maximum vibrational quantum number ($\lambda$).

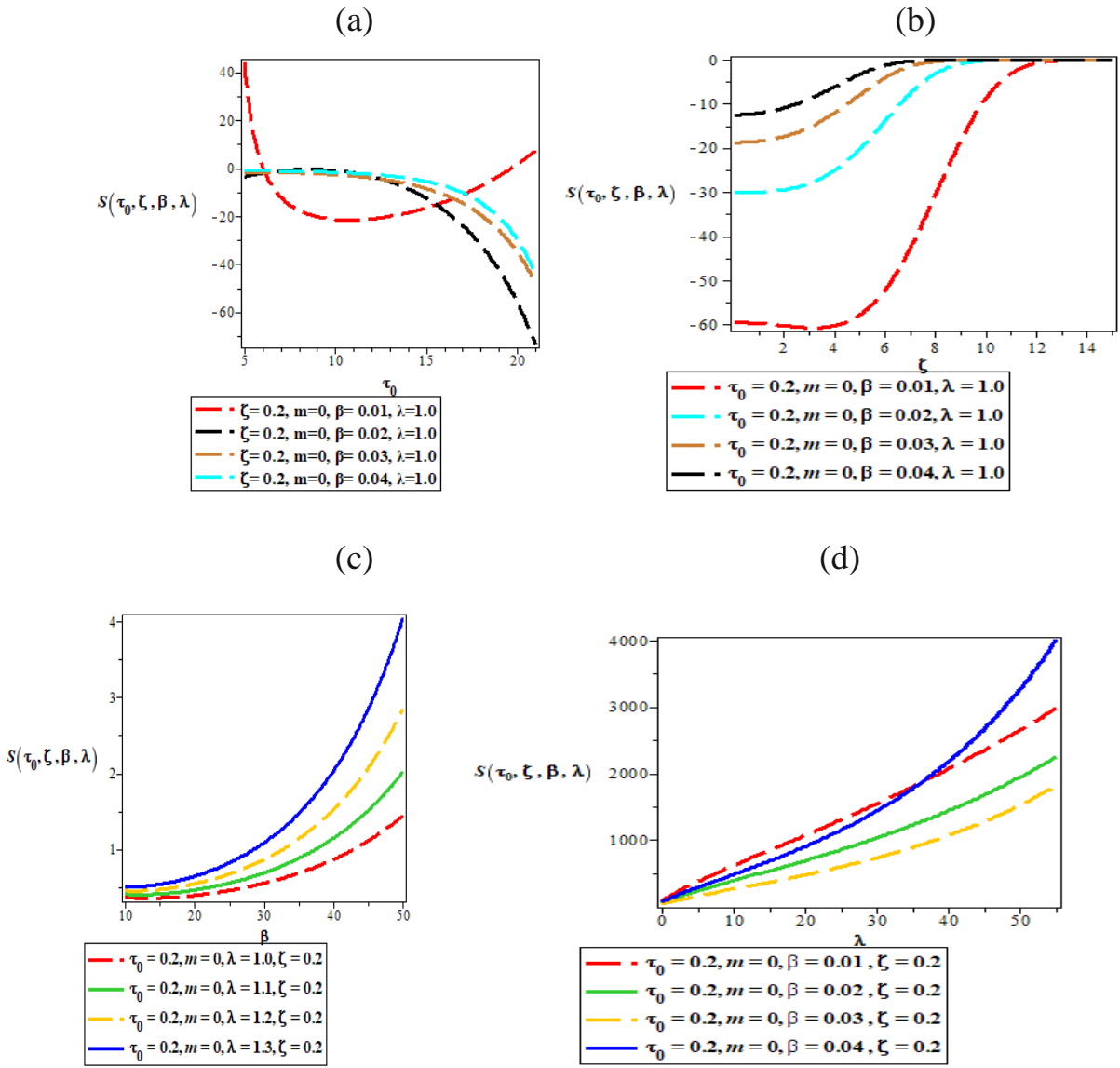

Figure 4: Variation of vibrational entropy with varying linear quantum flux with respect to (a) Magnetic flux ($\tau_0$), (b) Aharanov-Bohm flux ($\zeta$) (c) Inverse temperature parameter and (d) Maximum vibrational quantum number ($\lambda$).

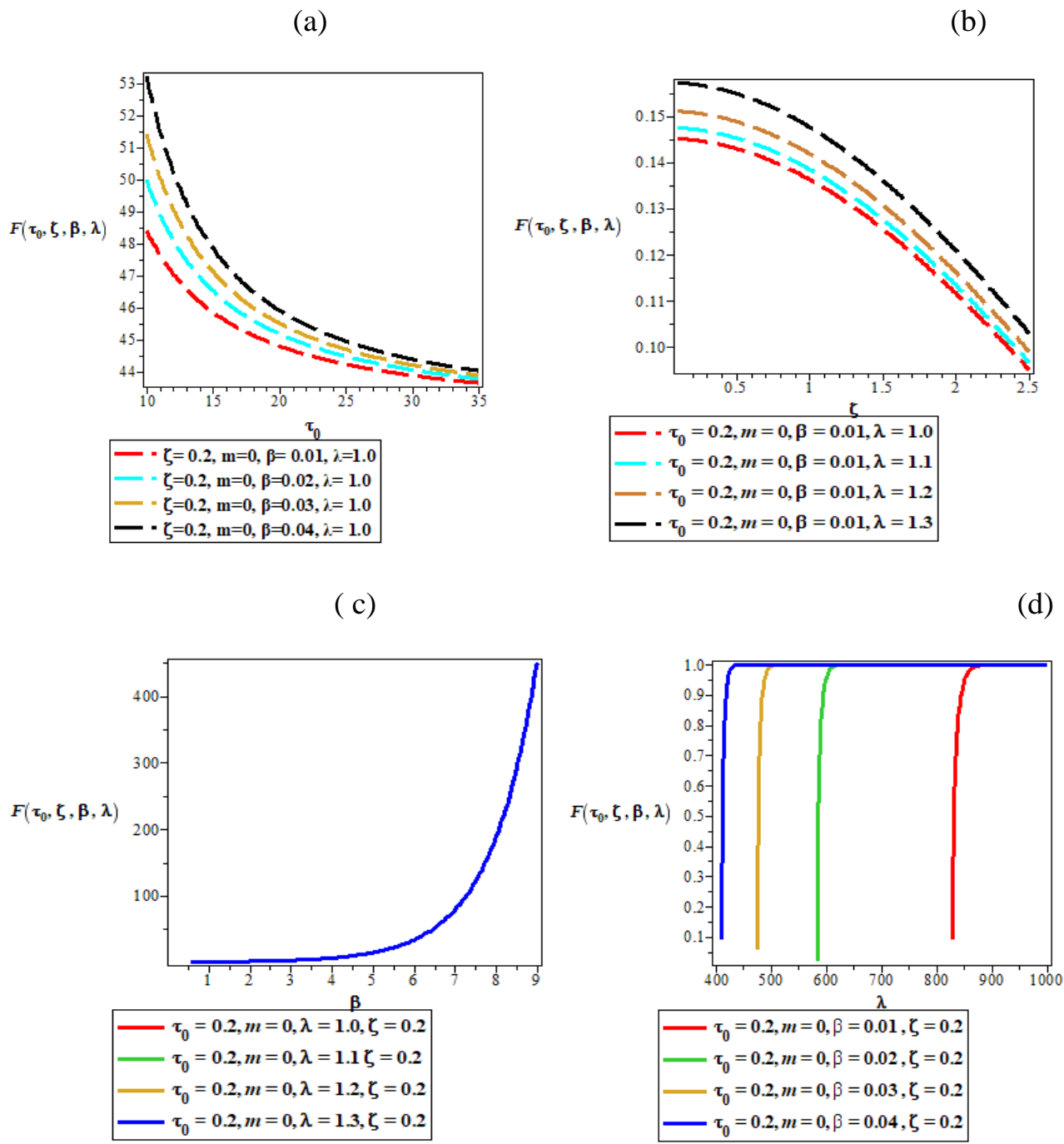

Figure 5: Variation of vibrational Free energy with varying linear quantum flux with respect to (a) Magnetic flux ($\tau_0$), (b) Aharanov-Bohm flux ($\zeta$) (c) Inverse temperature parameter and (d) Maximum vibrational quantum number ($\lambda$).

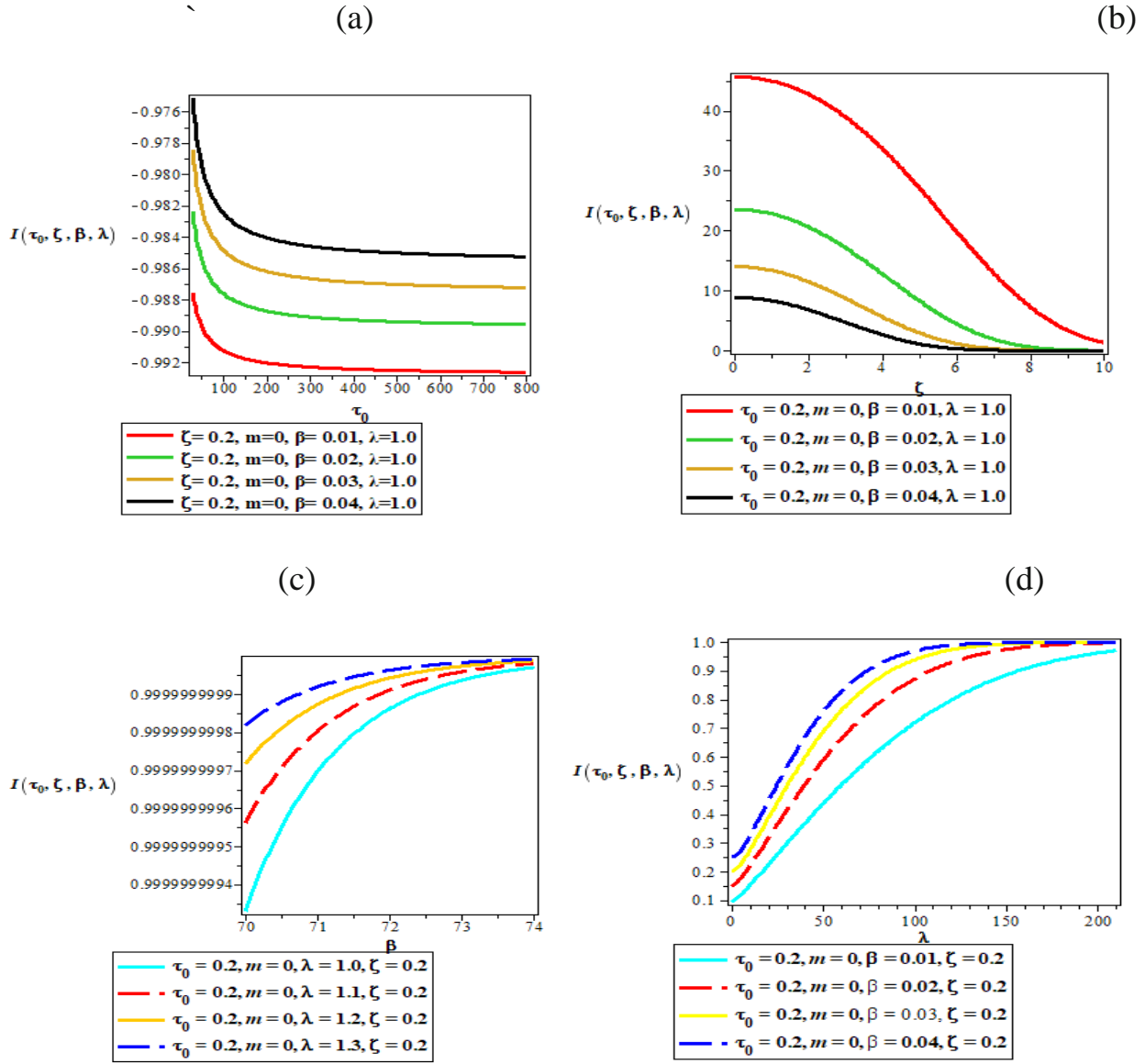

Figure 6: Variation of Persistent current with varying linear quantum flux with respect to (a) Magnetic flux ($\tau_0$), (b) Aharanov-Bohm flux ($\zeta$) (c) Inverse temperature parameter and (d) Maximum vibrational quantum number ($\lambda$).

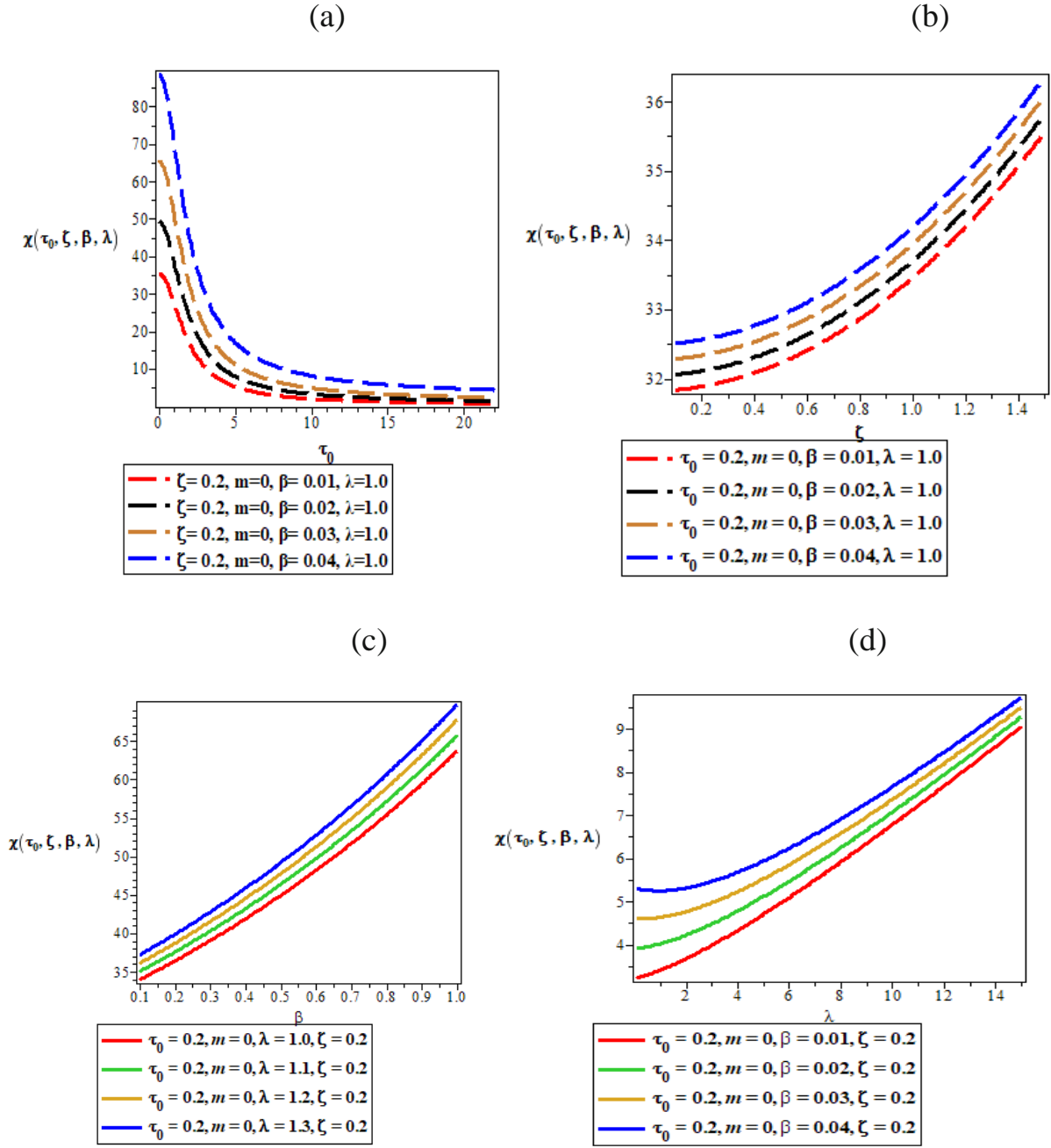

Figure 7: Variation of magnetic susceptibility with varying linear quantum flux with respect to (a) Magnetic flux ($\tau_0$), (b) Aharanov-Bohm flux ($\zeta$) (c) Inverse temperature parameter and (d) Maximum vibrational quantum number ($\lambda$).

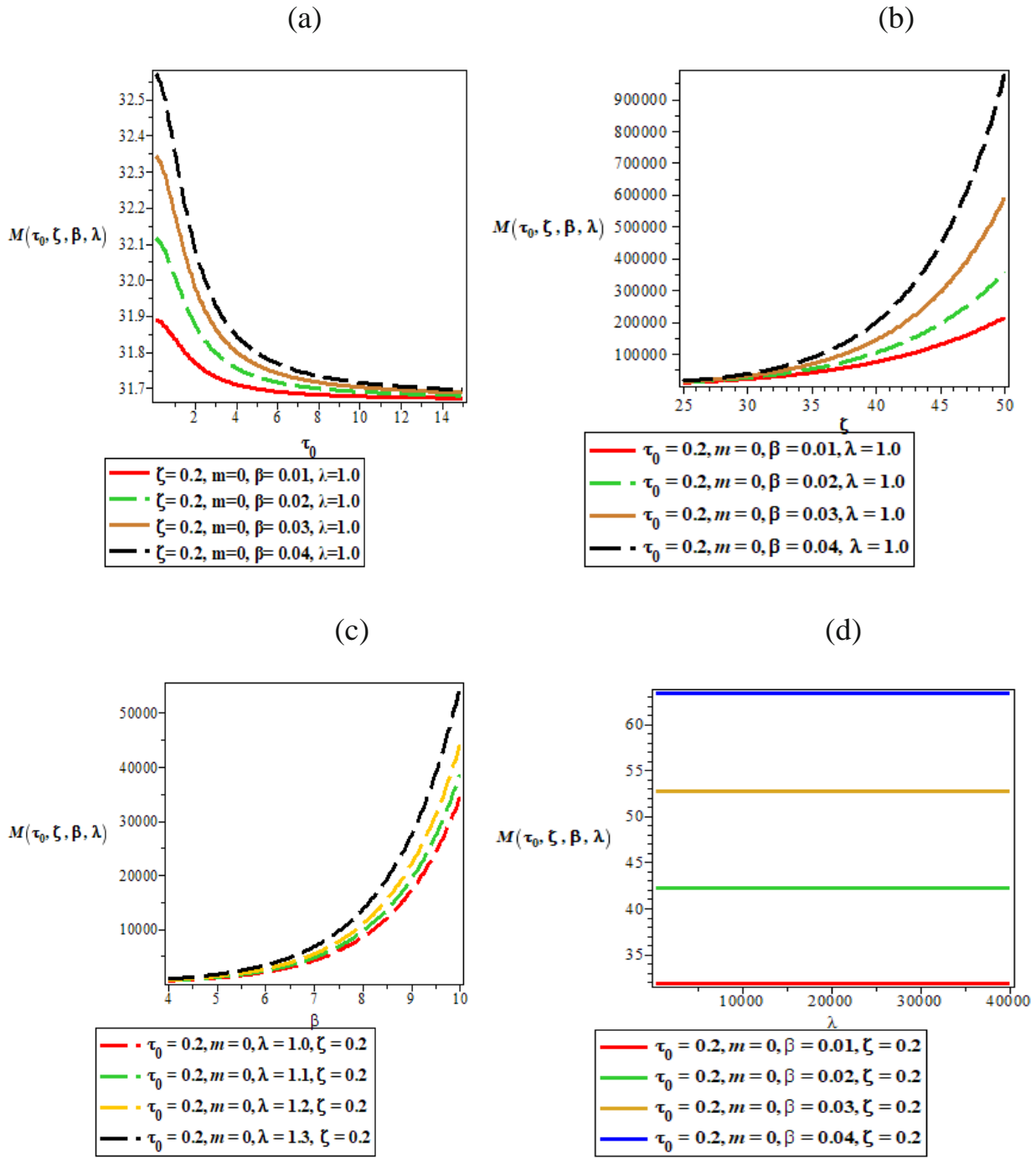

Figure 8: Variation of magnetisation with varying linear quantum flux with respect to (a) Magnetic flux ($\tau_0$), (b) Aharanov-Bohm flux ($\zeta$) (c) Inverse temperature parameter and (d) Maximum vibrational quantum number ($\lambda$).

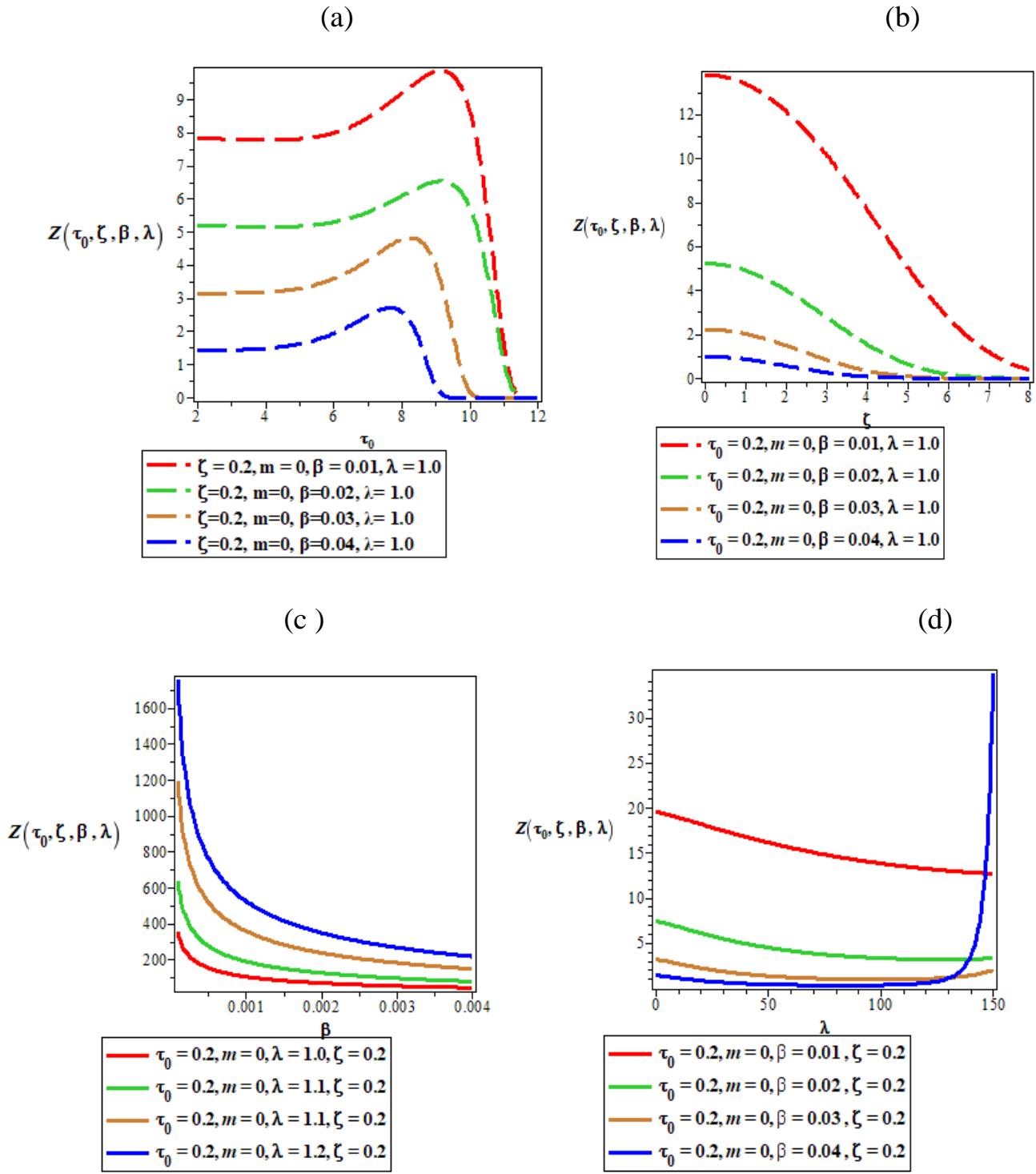

Figure 9: Variation of partition function with varying exponential quantum flux with respect to (a) Magnetic flux ($\tau_0$), (b) Aharanov-Bohm flux ($\zeta$) (c) Inverse temperature parameter and (d) Maximum vibrational quantum number ($\lambda$).

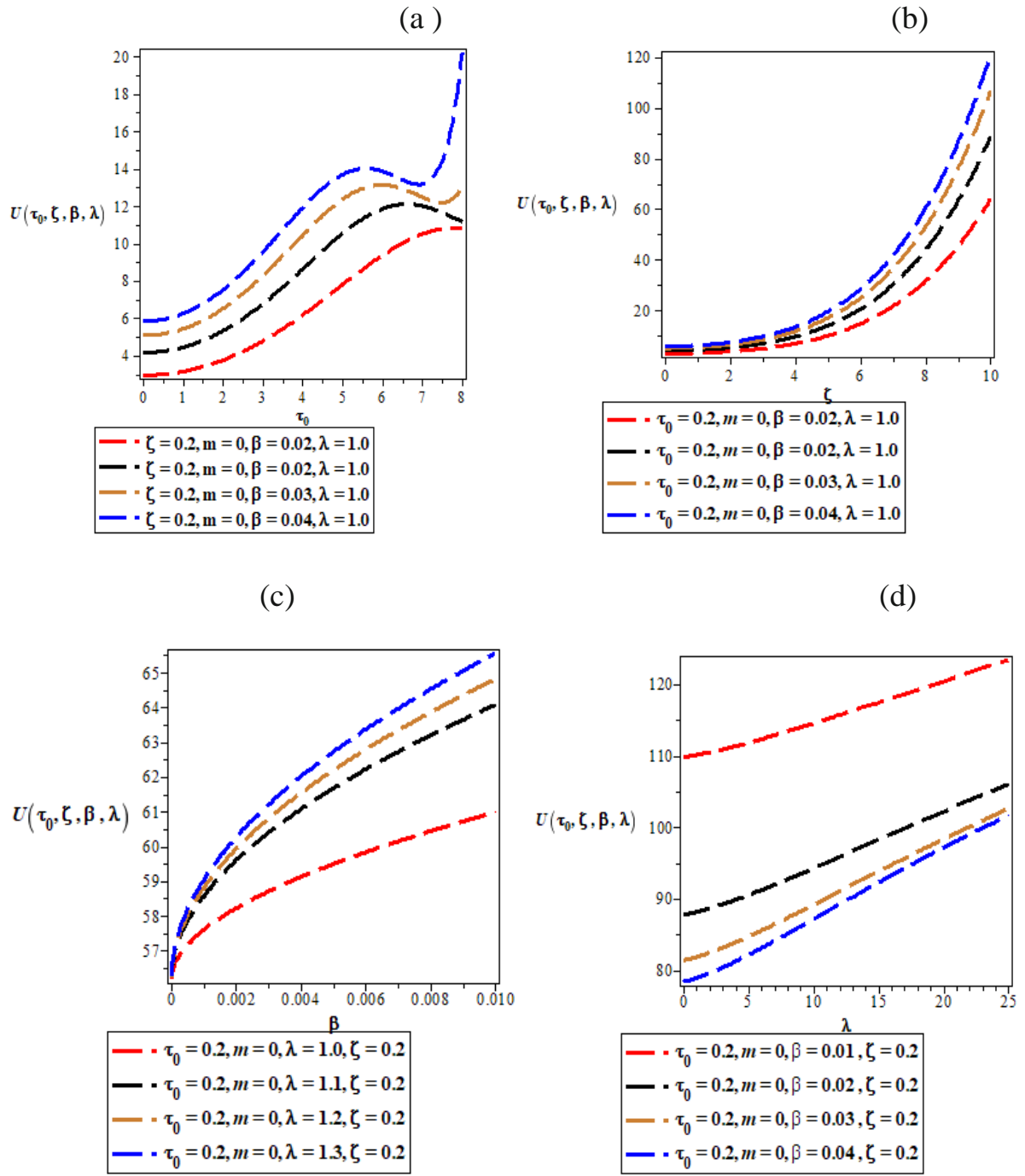

Figure 10: Variation of vibrational mean energy with varying exponential quantum flux with respect to (a) Magnetic flux ($\tau_0$), (b) Aharanov-Bohm flux ($\zeta$) (c) Inverse temperature parameter and (d) Maximum vibrational quantum number ($\lambda$).

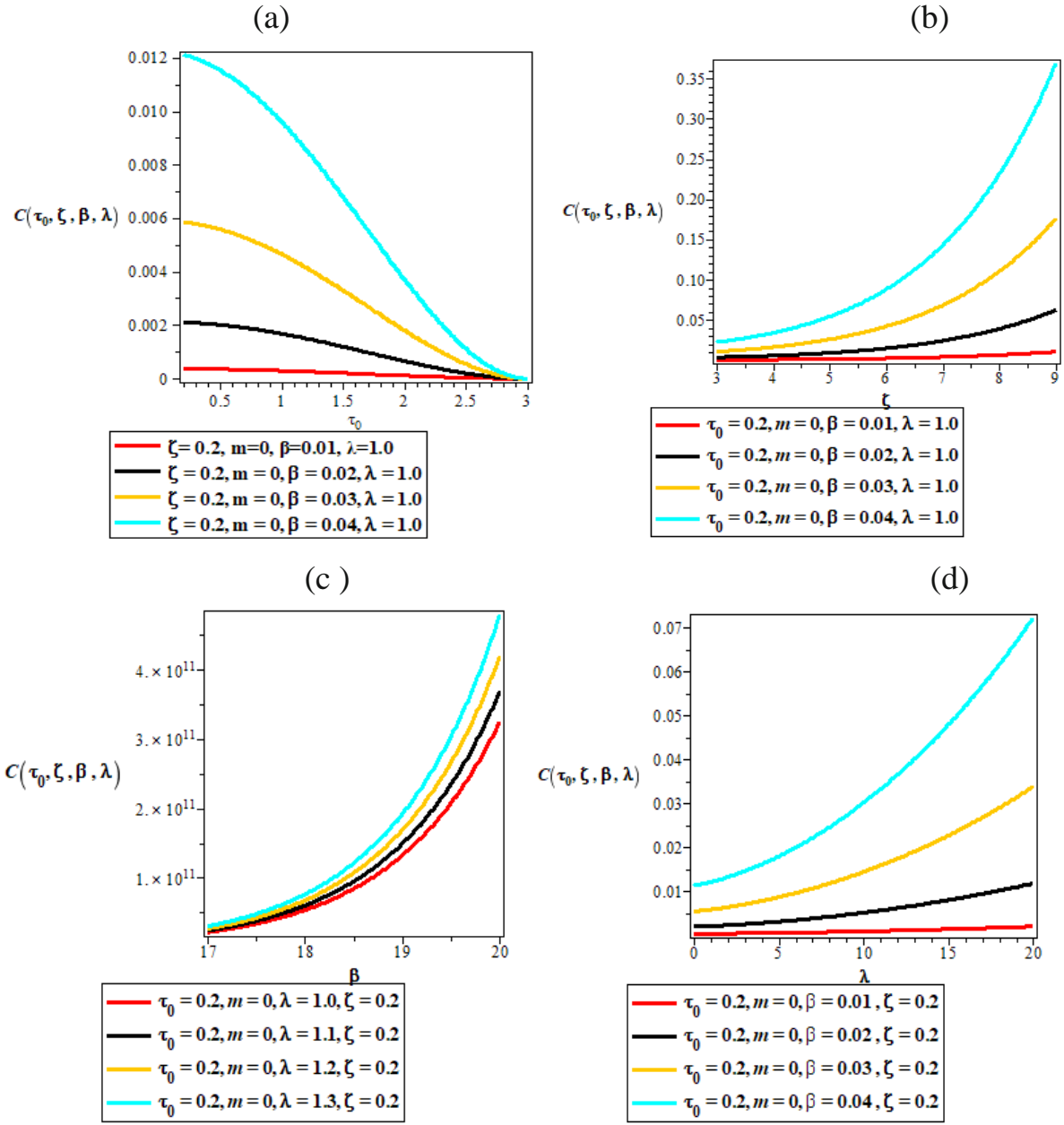

Figure 11: Variation of vibrational heat capacity with varying exponential quantum flux with respect to (a) Magnetic flux ($\tau_0$), (b) Aharanov-Bohm flux ($\zeta$) (c) Inverse temperature parameter and (d) Maximum vibrational quantum number ($\lambda$).

(a)            (b)

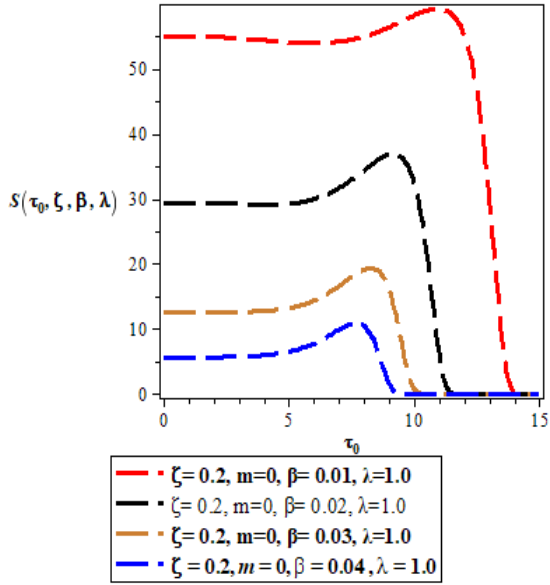
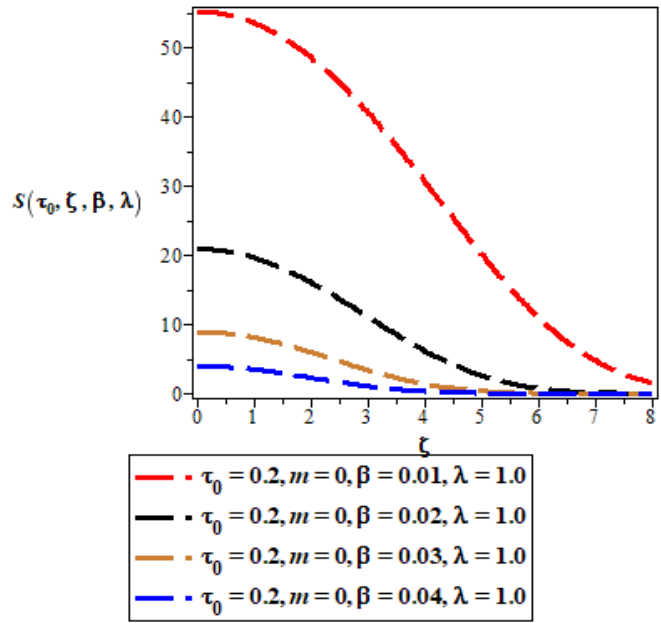

(c)                                                      (d)

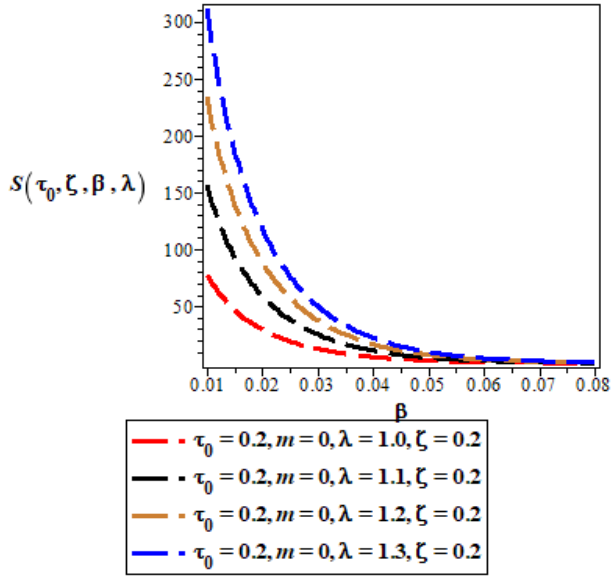
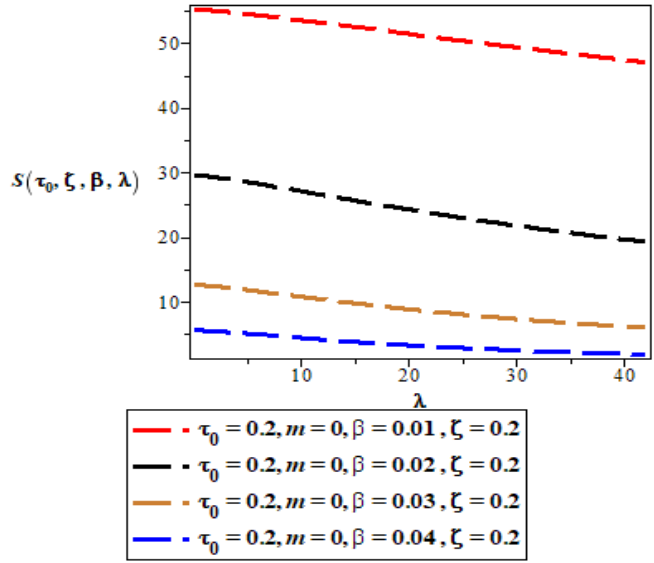

Figure 12: Variation of entropy with varying exponential quantum flux with respect to (a) Magnetic flux ($\tau_0$), (b) Aharanov-Bohm flux ($\zeta$) (c) Inverse temperature parameter and (d) Maximum vibrational quantum number ($\lambda$).

(a)                                                      (b)

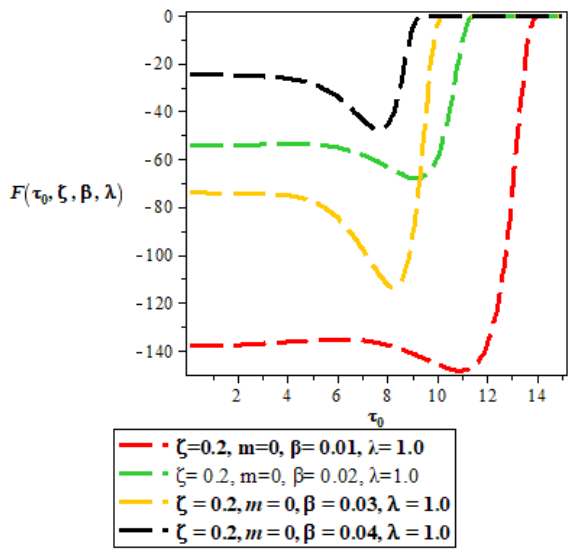

(c)

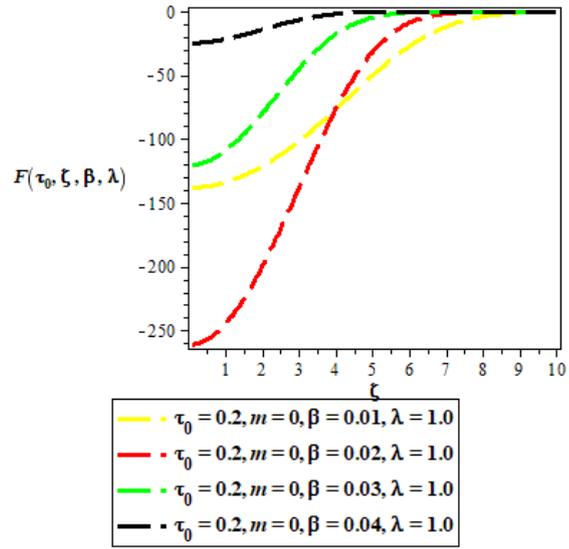

(d)

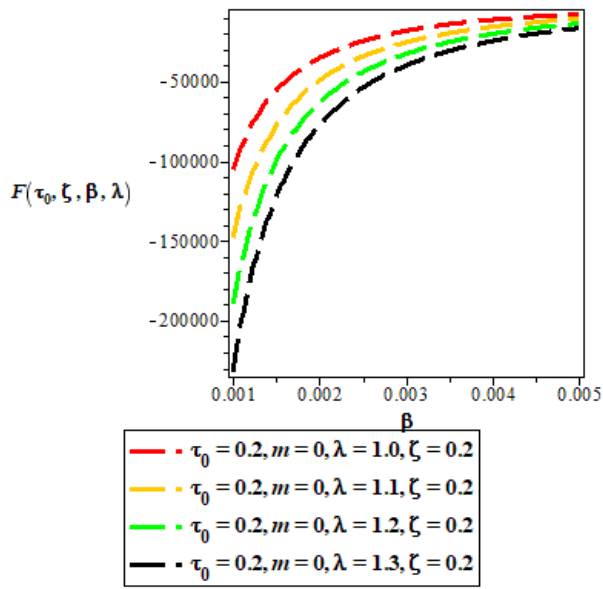

(a)

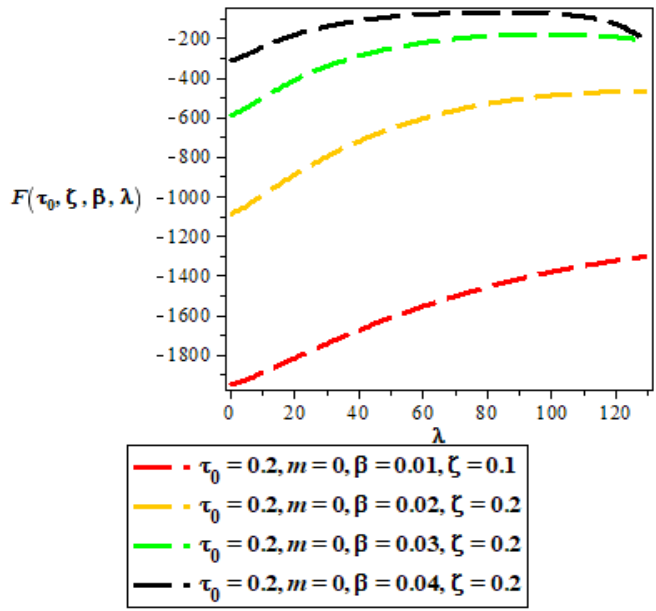

(b)

Figure 13: Variation of vibrational free energy with varying exponential quantum flux with respect to (a) Magnetic flux ($\tau_0$), (b) Aharanov-Bohm flux ($\zeta$) (c) Inverse temperature parameter and (d) Maximum vibrational quantum number ($\lambda$).

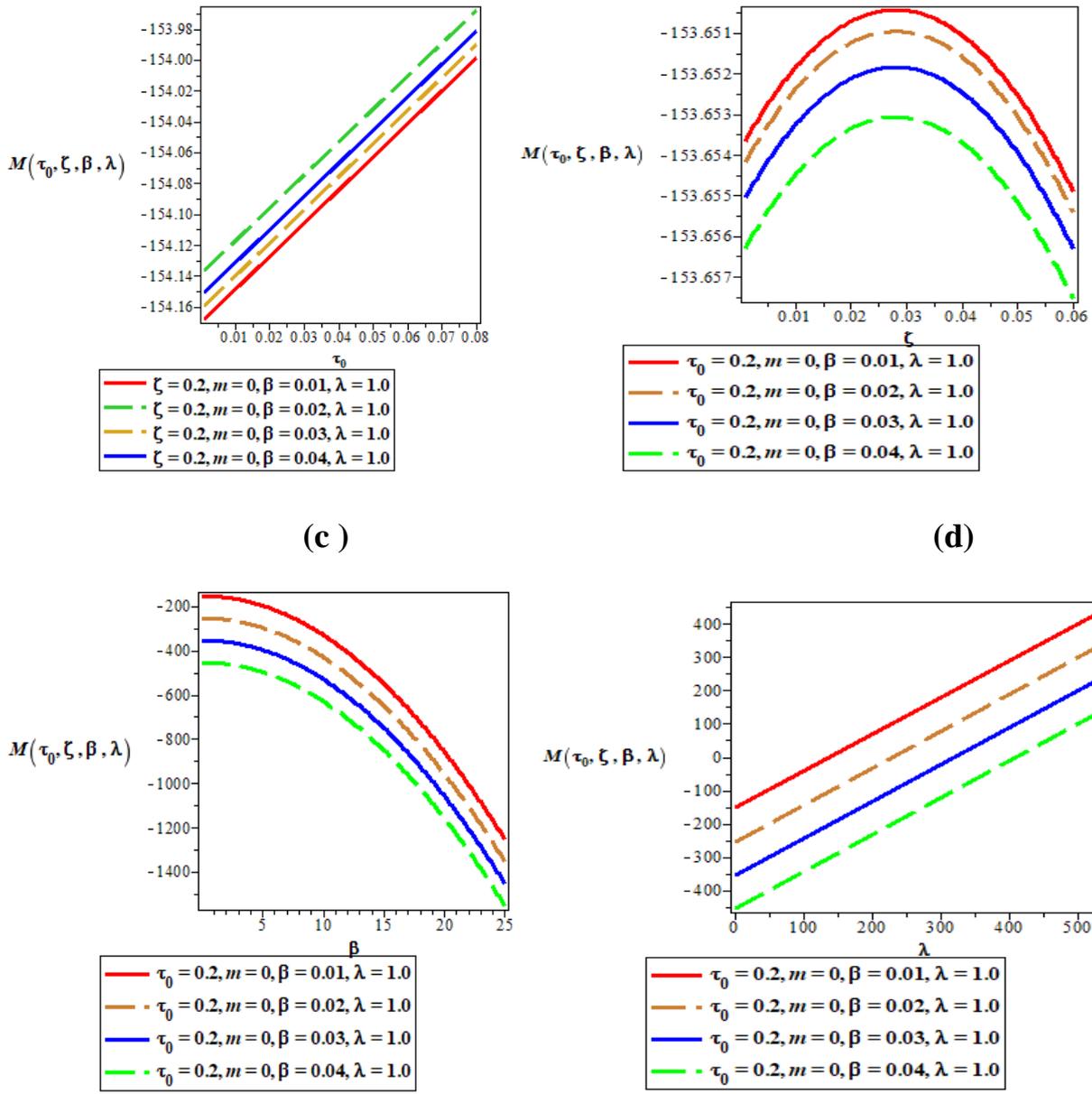

Figure 14: Variation of magnetisation with varying exponential quantum flux with respect to (a) Magnetic flux ($\tau_0$), (b) Aharanov-Bohm flux ($\zeta$) (c) Inverse temperature parameter and (d) Maximum vibrational quantum number ($\lambda$).

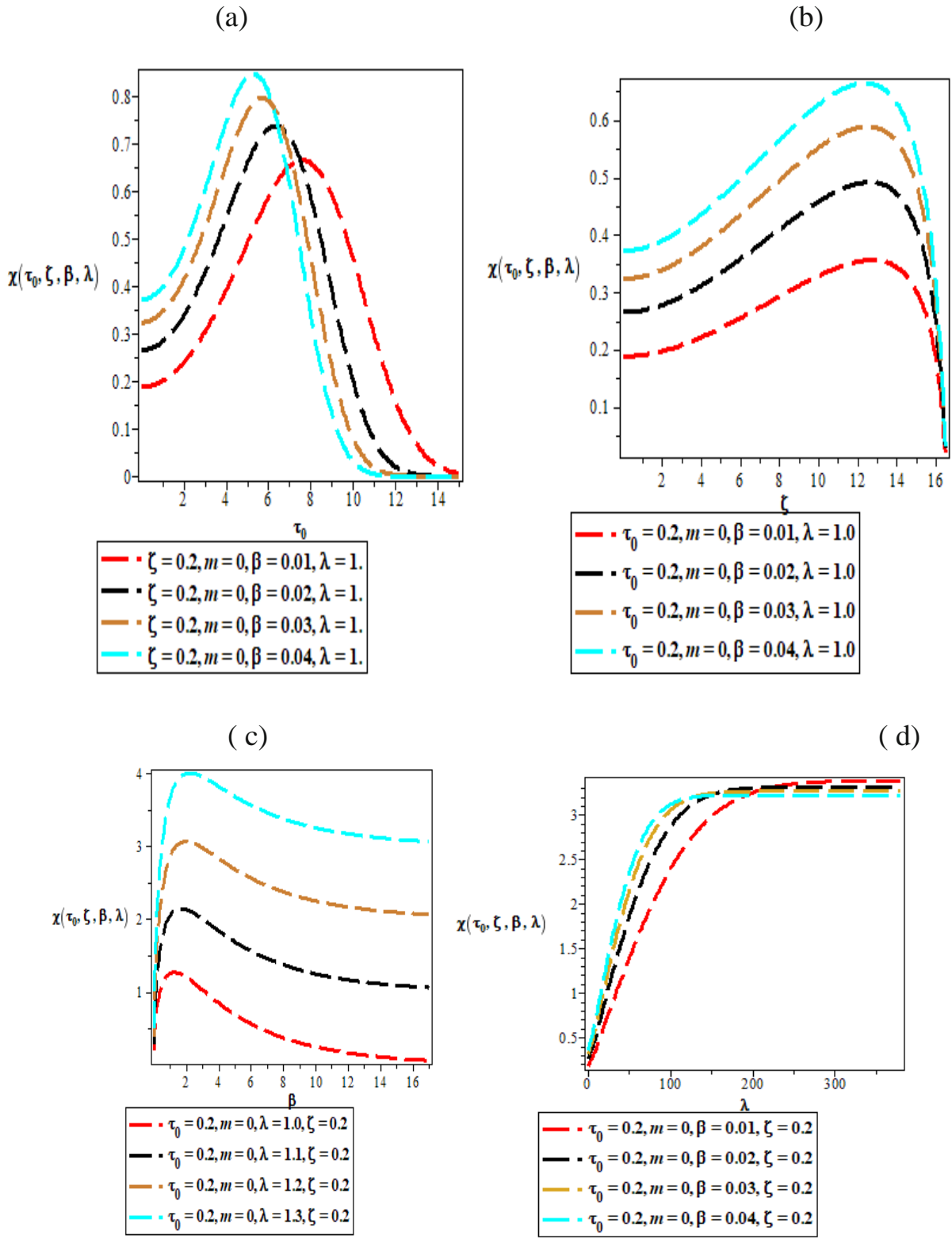

Figure 15: Variation of magnetic susceptibility with varying exponential quantum flux with respect to (a) Magnetic flux ($\tau_0$), (b) Aharanov-Bohm flux ($\zeta$) (c) Inverse temperature parameter and (d) Maximum vibrational quantum number ($\lambda$).

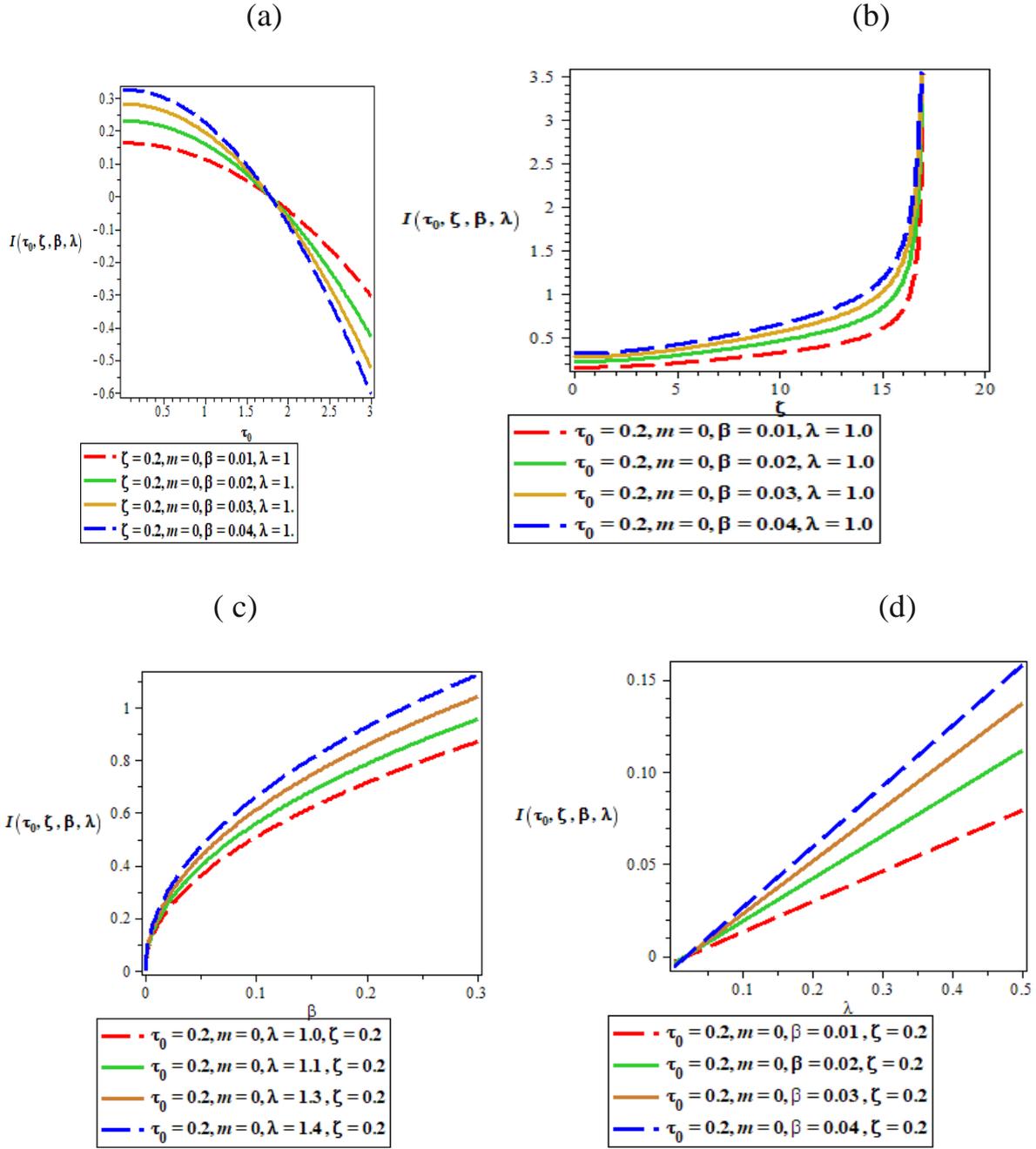

Figure 16: Variation of persistent current with varying exponential quantum flux with respect to (a) Magnetic flux ($\tau_0$), (b) Aharanov-Bohm flux ($\zeta$) (c) Inverse temperature parameter and (d) Maximum vibrational quantum number ($\lambda$).

## 6 Conclusion

In this work, we study 2D interacting Morse potential and its thermomagnetic properties using Nikiforov-Uvarov functional analysis method. We obtained the energy spectra for both linear and exponential varying quantum flux and applied to study partition function and other thermomagnetic properties. The numerical bound state solutions obtained for both linear and varying magnetic flux showcase degeneracies for some magnetic quantum spin. The thermos-magnetic properties has been well investigated for both cases.

APPENDIX

**Review of NU-Functional Analysis (NUFA) Method**

Ikot et al. [20] proposed a simple and elegant method for solving a second order differential equation of the hypergeometric type called Nikiforov-Uvarov-FunctionalAnalysis method (NUFA) method. This method is easy and simple just as the parametric NU method. As it is well-known the NU is used to solve a second-order differential equation of the form [16]

$$\psi''_n(s) + \frac{\tilde{\tau}(s)}{\sigma(s)}\psi'_n(s) + \frac{\tilde{\sigma}(s)}{\sigma^2(s)}\psi_n(s) = 0, \qquad (A1)$$

where $\sigma(s)$ and $\tilde{\sigma}(s)$ are polynomials, at most of second degree, and $\tilde{\tau}(s)$ is a first-degree polynomial. Tezcan and Sever [32] latter introduced the parametric form of NU method in the form

$$\psi'' + \frac{\alpha_1 - \alpha_2 s}{s(1-\alpha_3 s)}\psi' + \frac{1}{s^2(1-\alpha_3 s)^2}\left[-\xi_1 s^2 + \xi_2 s - \xi_3\right]\psi(s) = 0 \qquad (A2)$$

where $\alpha_i$ and $\xi_i (i=1,2,3)$ are all parameters. It can be observed in equation (A2) that the differential equation has two singularities at $s \to 0$ and $s \to \frac{1}{\alpha_3}$, thus we take the wave function in the form,

$$\psi(s) = s^\lambda (1-\alpha_3 s)^\nu f(s) \qquad (A3)$$

Substituting equation (A3) into equation (A2) leads to the following equation,

$$s(1-\alpha_3 s)f''(s)+\left[\alpha_1+2\mu-(2\mu\alpha_3+2\nu\alpha_3+\alpha_2)s\right]f'(s)$$

$$-\alpha_3\left(\mu+\nu+\frac{\alpha_2}{\alpha_3}-1+\sqrt{\frac{1}{4}\left(\frac{\alpha_2}{\alpha_3}-1\right)^2+\frac{\xi_1}{\alpha_3}}\right)\left(\mu+\nu+\frac{\alpha_2}{\alpha_3^2}-1+\sqrt{\frac{1}{4}\left(\frac{\alpha_2}{\alpha_3}-1\right)^2+\frac{\xi_1}{\alpha_3^2}}\right)$$

$$+\left[\frac{\mu(\mu-1)+\alpha_1\mu-\xi_3}{s}+\frac{\alpha_2\nu-\alpha_1\alpha_3\nu+\nu(\nu-1)\alpha_3-\frac{\xi_1}{\alpha_3}+\xi_2-\xi_3\alpha_3}{(1-\alpha_3 s)}\right]f(s)=0 \quad (A4)$$

Equation (A4) can be reduced to a Gauss hypergeometric equation if and only if the following functions vanished,

$$\lambda(\lambda-1)+\alpha_1\lambda-\xi_3=0 \quad (A5)$$

$$\alpha_2\nu-\alpha_1\alpha_3\nu+\nu(\nu-1)\alpha_3-\frac{\xi_1}{\alpha_3}+\xi_2-\xi_3\alpha=0 \quad (A6)$$

Thus, equation (A4) becomes

$$s(1-\alpha_3 s)f''(s)+\left[\alpha_1+2\mu-(2\mu\alpha_3+2\nu\alpha_3+\alpha_2)s\right]f'(s)$$

$$-\alpha_3\left(\mu+\nu+\frac{\alpha_2}{\alpha_3}-1+\sqrt{\frac{1}{4}\left(\frac{\alpha_2}{\alpha_3}-1\right)^2+\frac{\xi_1}{\alpha_3}}\right)\left(\mu+\nu+\frac{\alpha_2}{\alpha_3^2}-1+\sqrt{\frac{1}{4}\left(\frac{\alpha_2}{\alpha_3}-1\right)^2+\frac{\xi_1}{\alpha_3^2}}\right)f(s)=0 \quad (A7)$$

Solving equations (A5) and (A6) completely give,

$$\lambda=\frac{(1-\alpha_1)\pm\sqrt{(1-\alpha_1)^2+4\xi_3}}{2} \quad (A8)$$

$$\nu=\frac{(\alpha_3+\alpha_1\alpha_3-\alpha_2)\pm\sqrt{(\alpha_3+\alpha_1\alpha_3-\alpha_2)^2+4\left(\frac{\xi_1}{\alpha_3}+\alpha_3\xi_3-\xi_2\right)}}{2} \quad (A9)$$

Equation (A7) is the hypergeometric equation type of the form [20],

$$x(1-x)f''(x)+\left[\sigma+(\delta+g+1)x\right]f'(x)-\delta g\,f(x)=0 \quad (A10)$$

Using the quantization condition $\delta=-n$, together with equations (A9) and (A12), we obtain the energy equation and the corresponding wave equation respectively for the NUFA method as follows:

$$\mu^2+2\mu\left(\nu+\frac{1}{2}\left(\frac{\alpha_2}{\alpha_3}-1\right)+\frac{n}{\sqrt{\alpha_3}}\right)+\left(\nu+\frac{1}{2}\left(\frac{\alpha_2}{\alpha_3}-1\right)+\frac{n}{\sqrt{\alpha_3}}\right)^2-\frac{1}{4}\left(\frac{\alpha_2}{\alpha_3}-1\right)^2-\frac{\xi_1}{\alpha_3^2}=0 \quad (A11)$$

$$\psi(s)=Ns^{\frac{(1-\alpha_1)+\sqrt{(1-\alpha_1)^2+4\xi_3}}{2}}(1-\alpha_3 s)^{\frac{(\alpha_3+\alpha_1\alpha_3-\alpha_2)+\sqrt{(\alpha_3+\alpha_1\alpha_3-\alpha_2)^2+4\left(\frac{\xi_1}{\alpha_3^2}+\alpha_3\xi_3-\xi_2\right)}}{2}}{}_2F_1(\delta,g,\sigma;s) \quad (A12)$$

where $\delta$, $g$, $\sigma$ are given as follows,

$$\delta=\sqrt{\alpha_3}\left(\mu+\nu+\frac{1}{2}\left(\frac{\alpha_2}{\alpha_3}-1\right)+\sqrt{\frac{1}{4}\left(\frac{\alpha_2}{\alpha_3}-1\right)^2+\frac{\xi_1}{\alpha_3}}\right) \quad (A13)$$

$$g=\sqrt{\alpha_3}\left(\mu+\nu+\frac{1}{2}\left(\frac{\alpha_2}{\alpha_3}-1\right)-\sqrt{\frac{1}{4}\left(\frac{\alpha_2}{\alpha_3}-1\right)^2+\frac{\xi_1}{\alpha_3}}\right) \quad (A14)$$

$$\sigma = \alpha_1 + 2\mu \tag{A15}$$

The formulation is use to obtain the energy level and the corresponding wave function for the system. Consequently, the energy spectrum will be used to determine the thermodynamic properties for the system.